\documentclass[conference]{IEEEtran}
\IEEEoverridecommandlockouts
\usepackage{cite}
\usepackage{amsmath,amssymb,amsfonts}
\usepackage{algorithmic}
\usepackage{graphicx}
\usepackage{textcomp}
\usepackage{xcolor}
\usepackage{authblk}

\usepackage{subcaption}
\usepackage{stackengine}
\usepackage{amsfonts}
\usepackage{multirow}
\usepackage{wasysym}
\usepackage{color} 
\usepackage{soul}
\usepackage{listings}
\usepackage{mathtools}
\usepackage{cuted}
\usepackage[utf8x]{inputenc} 
\def\BibTeX{{\rm B\kern-.05em{\sc i\kern-.025em b}\kern-.08em
    T\kern-.1667em\lower.7ex\hbox{E}\kern-.125emX}}
\begin{document}

\title{Energy Efficient Algorithms based on VM Consolidation for Cloud Computing: \\ Comparisons and Evaluations
}

%

\author[1]{Qiheng Zhou}
\author[1]{Minxian Xu*\thanks{\IEEEauthorrefmark{1}Corresponding author: mx.xu@siat.ac.cn}}
\author[2]{Sukhpal Singh Gill}
\author[1]{Chengxi Gao} 
\author[3]{\\Wenhong Tian}
\author[1,4]{Chengzhong Xu}
\author[5]{Rajkumar Buyya}
\affil[1]{Shenzhen Institutes of Advanced Technology, Chinese Academy of Sciences, China}
\affil[2]{School of Electronic Engineering and Computer Science, Queen Mary University of London, UK}
\affil[3]{School of Software and Information Engineering, University of Electronic Science and Technology of China, China}
\affil[4]{Faculty of Science and Technology, Macao University, China}
\affil[5]{School of Computing and Information Systems, University of Melbourne, Australia}

\maketitle

\begin{abstract}
Cloud Computing paradigm has revolutionized IT industry and be able to offer computing as the fifth utility. With the pay-as-you-go model, cloud computing enables to offer the resources dynamically for customers anytime. Drawing the attention from both academia and industry, cloud computing is viewed as one of the backbones of the modern economy. However, the high energy consumption of cloud data centers contributes to high operational costs and carbon emission to the environment. Therefore, Green cloud computing is required to ensure energy efficiency and sustainability, which can be achieved via energy efficient techniques. One of the dominant approaches is to apply energy efficient algorithms to optimize resource usage and energy consumption. Currently, various virtual machine consolidation-based energy efficient algorithms have been proposed to reduce the energy of cloud computing environment. However, most of them are not compared comprehensively under the same scenario, and their performance is not evaluated with the same experimental settings. This makes users hard to select the appropriate algorithm for their objectives. To provide insights for existing energy efficient algorithms and help researchers to choose the most suitable algorithm, in this paper, we compare several state-of-the-art energy efficient algorithms in depth from multiple perspectives, including architecture, modelling and metrics. In addition, we also implement and evaluate these algorithms with the same experimental settings in CloudSim toolkit. The experimental results show the performance comparison of these algorithms with comprehensive results. Finally, detailed discussions of these algorithms are provided. 
\end{abstract}


\section{Introduction}
The emergence of cloud computing has contributed the computing resources as a new utility, like electricity and gas, and shaped the way how IT resources can be used by customers. Since its emergence, cloud computing has developed rapidly into one of the backbones of modern economy. Cloud consumers including government, research institutes and industry enterprises have all embraced and benefited from cloud computing significantly. Cloud computing also enables new business to be established within a short time, facilitates the expansion of enterprise globally, accelerates the progress of scientific research, and promotes the creation of various models and applications. Cloud service providers are also offering a variety of cloud services for customers with on-demand access to resources based on the pay-as-you-go model \cite{Buyya2018Manifesto}\cite{Kaur2015} \cite{GILL2019Trans}.

The infrastructure of cloud computing is cloud data centers. Currently, various cloud providers, including Amazon, Google, Microsoft, establish large-scale cloud data centers to fulfill the resources and services demands for customers. To ensure the availability and reliability, cloud data centers are required to be running 24/7. Nowadays, the majority of data centers spread of 300-4500 square meters containing hundreds to thousands of physical machines. A typical data center can consume up to 25,000 KWh per day. It is estimated that US data centers can consume 140 million kWh by 2020, which equals to 50 coal-based power plants. It is also claimed that the carbon footprint will reach 2-3\% of global emission \cite{Lavallee2014b}. To relieve the huge energy consumption and carbon emission, energy efficient techniques are required in cloud data centers \cite{Gill2019}\cite{Ghosh2018Adaptive}\cite{Soualah2017Energy}. 

Virtualization technology is an important part of green cloud data centers to support energy efficiency via Virtual Machine (VM) consolidation, where a VM is the software implementation of a computer that running with an operating system and applications \cite{Kaur2015}. VM consolidation is referred as the process the VMs can be reallocated from one physical machine to another without affecting the execution of users' requests. VM consolidation has been identified as one of the dominant energy efficient solutions to reduce the energy consumption of cloud data centers \cite{Gill2019}\cite{Xu2017a}. As VMs are packed on fewer physical machines via consolidation, idle physical machines can be turned off or switched to the low-power mode \cite{XuBrownoutSurvey}. 


VM consolidation has been proven to be an effective approach to reduce data center energy consumption, and various energy efficient algorithms based on VM consolidation have been proposed. These algorithms aim to optimize the placement of VMs to reduce the energy consumption while ensuring other constraints, e.g. Service Level Agreement (SLA) violations. Due to the uncontrolled network traffic, it is hard to conduct experiments under large-scale environment and reproduce results. Therefore, running experiments with validated simulation toolkit is an acceptable and reasonable way. Simulation toolkit can be easily applied to construct large-scale environment and generate reproducible results \cite{TIAN2015Open}. 

Among all existing cloud data centers simulation toolkits, CloudSim \cite{Calheiros2011a} is the most widely used one. CloudSim supports both system and behavior for cloud data centers, including physical machines, virtual machines, workloads and resource scheduling policies. The resource provisioning model is also generic so that it can be extended with ease and limited efforts. These attractive features have attracted users from hundreds of universities and research institutes, and some extended simulators complementing CloudSim have also been developed, e.g. CloudAnalyst \cite{wickremasinghe2010cloudanalyst} and NetworkCloudSim \cite{garg2011networkcloudsim}. 



The motivation of this article is to analyze the problem of rising energy consumption in depth through a comparison of some state-of-the-art energy efficient algorithms in cloud data centers. This article is particularly motivated from the following facts:
\begin{itemize}
	\item The need and demand for understanding existing VM-based energy efficient algorithms for cloud data centers.
	\item The proposed algorithms were evaluated in different scenarios and configurations, thus their advantage and disadvantage are not carefully examined. 
	\item The requirement for selecting the best suitable algorithm based on different priorities. 
\end{itemize}

In this paper, we select various well-known VM consolidation-based energy efficient algorithms and implement experiments using CloudSim toolkit. 
The evaluated algorithms are picked from state-of-the-art energy efficient algorithms that have shown good performance in energy efficiency. 

The main \textbf{contributions} of this work are as follows:
\begin{itemize}
	\item Offering a cross-sectional view of the investigated VM consolidation-based energy efficient algorithms, which present outstanding performance in cloud computing area.
	
	{\color{black}\item Presenting a unified simulation-based analysis framework based on CloudSim that allows evaluation and comparison of energy-efficient VM consolidation algorithms in a unified and unbiased way.}
	
	\item Discussing the advantages and disadvantages of the investigated algorithms to provide suggested algorithms for different scenarios.
\end{itemize}

The rest of the paper is organized as: We provide an overview on VM consolidation-based energy efficient algorithm for cloud computing in Section 2. The investigated algorithms are introduced in Section 3. Section 4 discusses the modelling of the investigated algorithms, and the adopted metrics in the investigated algorithms are summarized in Section 5. Section 6 shows the performance comparison of the investigated algorithms. Finally, conclusions and future research trends are given.

\section{Related Work}

\subsection{Energy Efficient Algorithms in Clouds}
A few articles have conducted surveys or taxonomies on energy efficient algorithms based on VM consolidation for cloud data centers. Mansouri  et al. \cite{Mansouri2017} introduced a survey on resource management in cloud environment, and discussed VM consolidation-based energy efficient algorithms from cloud management system level. Kaur et al. \cite{Kaur2015} presented a comprehensive survey for energy efficient scheduling approach in clouds, and compared some consolidation-based techniques, including VM consolidation-based approach. Orgerie et al. \cite{Orgerie2014} conducted a survey on techniques for improving the energy efficiency for large-scale distributed systems, and surveyed on VM migration algorithms for Clouds. Gill et al. \cite{Gill2019} proposed a taxonomy for sustainable cloud computing and introduced a taxonomy for VM consolidation-based algorithms without focusing on energy efficiency. Mann et al. \cite{mann2015allocation} introduced a survey on VM allocation in cloud data centers from problem modelling and optimization algorithms perspectives. Ahmad et al. \cite{ahmad2015survey} conducted a survey on VM migration and server consolidation framework for cloud data centers, in which the commonalities and differences of investigated VM migration algorithms are highlighted. 

However, these surveys and taxonomy focused on the comparison with high-level comparison of VM consolidation-based energy efficient algorithms without evaluating the performance under experimental environments. Thus, our work advances the previous works by evaluating the state-of-the-art algorithms not only from modelling perspective but also with experimental comparisons. It also identifies the merits and demerits of investigated algorithms to provide suggestions for research in related areas.

\subsection{VM consolidation-based Energy Efficient Algorithms}
Beloglazov et al. \cite{beloglazov2012energy} proposed an energy-aware allocation algorithm of data center resources that focuses on VM scheduling named Modified Best Fit Decreasing (MBFD). Their objective is to reduce the energy consumption of data centers while ensuring SLA. Mastroianni et al. \cite{mastroianni2013probabilistic} proposed energy efficient scheduling policy based on probabilistic procedures. Li et al. \cite{li2017holistic} presented a holistic VM scheduling algorithm capable of minimizing total data center energy consumption, including computing energy and cooling energy.  Ranjbari et al. \cite{ranjbari2018learning} introduced an algorithm based on learning automata for energy and SLA efficient consolidation of VMs in cloud data centers. Farahnkian et al. \cite{ACW2015} proposed a novel dynamic VM consolidation approach based on Ant Colony Optimization to reduce the energy consumption of data centers. 

In the performance evaluation of these algorithms, MBFD has been evaluated as the baseline, however, these algorithms are not compared and evaluated together, thus it is hard to identify the performance of these VM consolidation-based energy efficient algorithms. In this paper, we aim to compare these algorithms in depth and evaluate them under the same configurations to show the performance comparison.

\section{Overview of the investigated algorithms}
 \begin{figure}[!ht]
	\centering
	\includegraphics[width=0.95\linewidth]{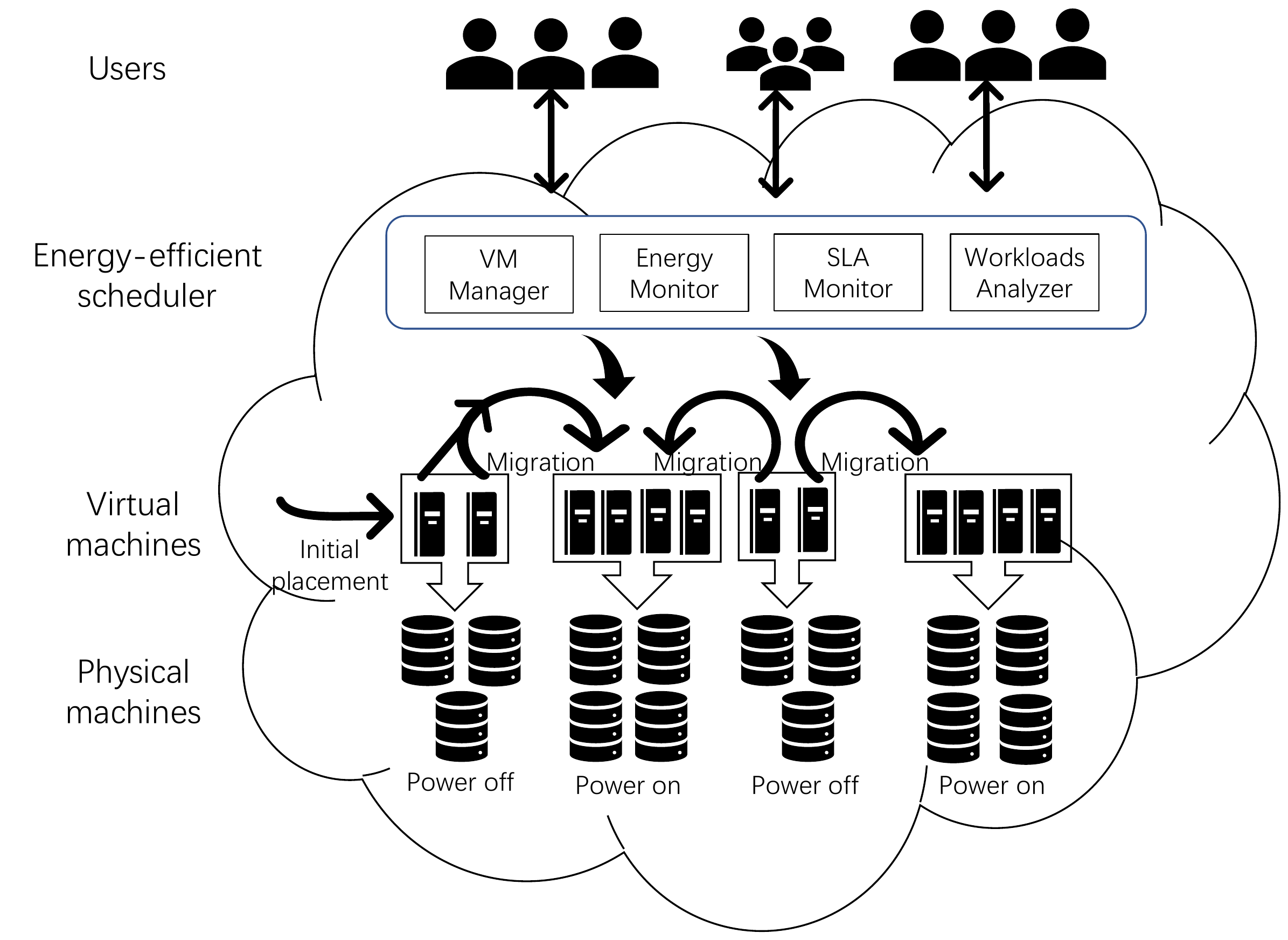}
	
	\caption[VarPerOptCom]{Energy efficient scheduling based on VM consolidation}

	\label{fig:vmconsolidation}
\end{figure}

Fig. \ref{fig:vmconsolidation} shows the high-level architecture of VM consolidation-based energy efficient algorithms in cloud data centers. This architecture is generic and our investigated algorithms all conform it. The architecture mainly has 4 entities: 

\textit{1. Users:} Cloud users submit their requests to cloud, and the requests will be processed by cloud services. 

\textit{2. Energy efficient scheduler:} It can contain the following components to support the energy efficient scheduling and act as the interface between users and provisioned resources from service providers:

\textit{(a) VM manager:} This component tracks the resource usage of VMs and makes decisions on VMs behaviors, including when and where to consolidate VMs. To achieve this objective, it requires the energy and SLA information from Energy Monitor and SLA monitor.

\textit{(b) Energy Monitor:} It monitors the energy usage in the system caused by VMs and physical machines. It also provides information for the decisions of the VM manager. 

\textit{(c) SLA Monitor:} It monitors how SLA is influenced by the operations in the system. It can also represent performance constraints when reducing energy consumption in the system.  

\textit{(d) Workloads Analyzer:} This component interprets users' requests and service requirements before processing them. The workloads can be distributed to different VMs based on resource characteristics.

\textit{3. Virtual machines:} The workloads are deployed and executed on VMs. The VMs can be managed according to the incoming workloads with two phases: initial placement and VM migration. VMs are placed by initial placement algorithm to be allocated to physical machines initially. According to workloads, the placement can be optimized via VM consolidation algorithm, thus the unused machines can be temporarily turned off or switched to low-power mode. 

\textit{4. Physical machines:} The infrastructure offers physical machines to provision virtualized resources to meet users' requests.

 Based on the following criteria, we carefully select 5 state-of-the-art VM consolidation-based energy efficient algorithms for our comparisons and evaluations:
\begin{itemize}
	\item To make the comparison more persuasive, the algorithms were published in prominent journals or conferences, and the algorithms can be representative of a category of algorithms.
	\item To ensure the evaluation results reproducible, the algorithms were implemented in CloudSim or can be easily evaluated in CloudSim.
	\item To make the algorithm comparable, the algorithms should have been evaluated with the same baseline.
\end{itemize}

In the following subsections, we will provide the overview of our investigated algorithms. Table \ref{tab:comparison} shows the comparison of the investigated algorithms based on different parameters such as application type, operational environment, objective function, scheduling mechanism, scheduling criteria, type of workloads and identified merits and demerits of each algorithm.  

\begin{table*}[!ht]
	\caption{High-level Comparison of Investigated Algorithms}
	\label{tab:comparison}
	\resizebox{\textwidth}{!}{%
		\begin{tabular}{|c|c|c|c|c|c|c|c|c|c|}
			\hline
			\textbf{Algorithm} & \textbf{\begin{tabular}[c]{@{}c@{}}Application\\ Type\end{tabular}} & \textbf{\begin{tabular}[c]{@{}c@{}}Operational\\ Environment\end{tabular}} & \textbf{\begin{tabular}[c]{@{}c@{}}Objective\\ Function\end{tabular}} & \textbf{\begin{tabular}[c]{@{}c@{}}Power\\ Model\end{tabular}} & \textbf{\begin{tabular}[c]{@{}c@{}}Power\\ Component\end{tabular}} & \textbf{\begin{tabular}[c]{@{}c@{}}Scheduling\\ Mechanism\end{tabular}} & \textbf{Workloads} & \textbf{Merits} & \textbf{Demerits} \\ \hline
			MBFD\cite{beloglazov2012energy} & \begin{tabular}[c]{@{}c@{}}Dynamic\\ workloads \\ (web service)\end{tabular} & \begin{tabular}[c]{@{}c@{}}Distributed\\ and\\ Heterogeneous\end{tabular} & \begin{tabular}[c]{@{}c@{}}To optimize\\ energy\\ consumption\end{tabular} &  Linear & \begin{tabular}[c]{@{}c@{}}CPU and\\ memory\end{tabular} & \begin{tabular}[c]{@{}c@{}}Dynamic\\ Consolidation\\ (Proactive)\end{tabular} & \begin{tabular}[c]{@{}c@{}}Heterogeneous\\ workloads\end{tabular} & \begin{tabular}[c]{@{}c@{}}Reduced energy\\ consumption and\\ SLA violation rate\end{tabular} & \begin{tabular}[c]{@{}c@{}}There is a need\\ of holistic resource\\ management\end{tabular} \\ \hline
			
			EcoCloud \cite{mastroianni2013probabilistic}& \begin{tabular}[c]{@{}c@{}}Dynamic\\ workloads\end{tabular} & \begin{tabular}[c]{@{}c@{}}Distributed \\ and homogeneous\end{tabular} & \begin{tabular}[c]{@{}c@{}}To improve consolidation\\ of VMs\end{tabular} &  Linear & \begin{tabular}[c]{@{}c@{}}Memory and\\ CPU\end{tabular} & \begin{tabular}[c]{@{}c@{}}Bounouli-based \\ scheduling\\
				(Proactive)\end{tabular} & \begin{tabular}[c]{@{}c@{}}Planetlab workloads\\ traces\end{tabular} & \begin{tabular}[c]{@{}c@{}}Evaluated the \\ effect of power\\ consumption on SLA\\ violations\end{tabular} & \begin{tabular}[c]{@{}c@{}}To investigate the impact\\ of number of VM migrations on\\ SLA violations\end{tabular} \\ \hline
			
			GRANITE \cite{li2017holistic}& Heterogeneous & Distributed & \begin{tabular}[c]{@{}c@{}}To investigate the \\ temperature distribution\\ of airflow and server CPU\end{tabular}  & Linear & \begin{tabular}[c]{@{}c@{}}Cooling, \\ storage, memory,\\ CPU and network\end{tabular} & \begin{tabular}[c]{@{}c@{}}2D-Computational Fluid\\ Dynamics (CFD) modelling-\\ based scheduling \\ (Proactive)\end{tabular} & \begin{tabular}[c]{@{}c@{}}Google datacenter\\ trace logs\end{tabular} & \begin{tabular}[c]{@{}c@{}}Minimizing total\\ datacenter energy\\ consumption (cooling\\ and computing)\end{tabular} & \begin{tabular}[c]{@{}c@{}}To improve accuracy, \\ 2-dimensional CFD model can be\\ enhanced to 3-dimensional \\ CFD model\end{tabular} \\ \hline
			
			LOAD \cite{ranjbari2018learning}& \begin{tabular}[c]{@{}c@{}}Dynamic \\ workloads\end{tabular} & Homogeneous & \begin{tabular}[c]{@{}c@{}}To optimize energy\\ consumption, number of \\ VM migrations and SLA\\ violations\end{tabular}  & Linear & CPU & \begin{tabular}[c]{@{}c@{}}Learning automata \\ based scheduling \\ (proactive)\end{tabular} & \begin{tabular}[c]{@{}c@{}}CPU\\ utilization\end{tabular} & \begin{tabular}[c]{@{}c@{}}Improved CPU\\ utilization and reduced\\ SLA violations and\\ energy consumption\end{tabular} & \begin{tabular}[c]{@{}c@{}}Under-utilization of \\ resource is not \\ considered\end{tabular} \\ \hline
			
			ACS \cite{ACW2015} & \begin{tabular}[c]{@{}c@{}}CPU and\\ memory-intensive\\ workloads\end{tabular} & \begin{tabular}[c]{@{}c@{}}Distributed\\ and\\ Heterogeneous\end{tabular} & \begin{tabular}[c]{@{}c@{}}To investigate the\\ relationship between\\ energy consumption, VM\\ migrations and QoS\end{tabular}  & Linear & \begin{tabular}[c]{@{}c@{}}CPU and,\\ Memory \end{tabular} & \begin{tabular}[c]{@{}c@{}}Ant colony \\ optimization and \\ (reactive)\end{tabular} & \begin{tabular}[c]{@{}c@{}}Heterogeneous\\ workloads\end{tabular} & \begin{tabular}[c]{@{}c@{}}Reduced energy and\\ VM migrations\end{tabular} & \begin{tabular}[c]{@{}c@{}}The impact of VM migration \\ on network bandwidth can be \\ investigated to reduce power\\ consumption further\end{tabular} \\ \hline
			

		\end{tabular}%
	}
\end{table*}

\subsection{MBFD}
Modified Best Fit Decreasing (MBFD) \cite{beloglazov2012energy} aims to reduce the energy consumption of data centers while ensuring SLA. It solves the VM initial placement phase by regarding it as a bin packing problem. MBFD is designed to assign the VMs to the hosts that produce the minimum incrementation of energy consumption. In the VM consolidation phase, the algorithm optimizes the VM allocation via consolidation for better energy efficiency, the target host is also selected as in the initial placement. The motivation of the proposed algorithm is 
due to the uncertainty of task lifecycles, some VMs are likely to host an over-provisioned of applications while others run with low resource utilization. Unbalanced workloads in cloud data centers result in a severe waste of resources and performance degradation. The VM consolidation in this work is a proactive process and can be applied with heterogeneous workloads.  

MBFD has been evaluated as the baseline for many VM consolidation-based energy efficient algorithm. Various recent algorithms were proposed to improve the performance of this algorithm. The advantage of this algorithm is that it considers the trade-offs between the reduced energy and SLA violation and it is easy to implement. The limitation is that it does not consider the holistic resource management, which is complemented by recent research.

\subsection{EcoCloud}
EcoCloud \cite{mastroianni2013probabilistic} is an energy efficient scheduling policy based on probabilistic procedures. VM assignment and VM migration are the two phases in the algorithm. In the VM assignment phase, different from the MBFD introduced in the previous section, the VM manager sends an invitation to a subset of all active hosts to obtain a list of hosts that accept the incoming VM. After receiving the invitation, the host performs a Bernoulli trial, which computes the value of an overall assignment function to decide whether to accept a VM or not. The assignment function considers the resource utilization and utilization threshold, then calculates the probability to accept the incoming VM. If the trial succeeds, the host sends back a message to the manager. The manager collects all messages and selects one available host to allocate the incoming VM. The second part of EcoCloud is VM migration, which utilizes two Bernoulli trial based functions to handle the over-utilized and under-utilized situations separately. If the trial of the under-utilized situation is successful, a randomly selected VM will be migrated. On the other hand, migrations with over-utilized will migrate the VM that decreases resource utilization to be lower than the over-utilized threshold. The target host to accept the VM is decided by another assignment function with a slight modification of a parameter. 

EcoCloud can be the representative algorithm aiming to reduce energy while minimizing VM migrations. The main advantage of EcoCloud is that the approach can reduce downtime duration and transmission bandwidth. Furthermore, hosts can make migration decisions by themselves, thus the pressure of VM manager is relieved. The limitation of this work is that it investigates the effect of power consumption on SLAs while the effects of VM migrations on SLAs are not evaluated. 

\subsection{GRANITE}
GRANITE \cite{li2017holistic} is a holistic virtual machine scheduling algorithm capable of minimizing total data center energy consumption, including computing energy and cooling energy. This work considers Computer Room Air Conditioner (CRAC) as the only cooling devices and constructed their models. Based on server models and cooling models, GRANITE utilizes a greedy algorithm to conduct the VM initial placement and dynamic migration. They assume that the users' resource demand can be predicted. In the initial placement stage, the greedy algorithm is applied in GRANITE for all VMs to select the host with the least increment of total energy after the placement. The CRAC will be adjusted if the CPU temperature is above the threshold. In the dynamic VM consolidation stage, the algorithm aims to balance the workloads and cooling energy consumption. The GRANITE defines a dynamic temperature threshold and checks host status. If the host temperature is above the temperature threshold, a set of VMs will be migrated to other hosts. The target host selection for migration is the same as the greedy algorithm in the initial placement stage. 

GRANITE can present the algorithms that consider the holistic management of energy in cloud data centers. The core idea of the algorithm is similar to MBFD algorithm, while it considers cooling power, which leads to more accurate and holistic scheduling results. The advantage of this work is that it combines the server status and data center thermal control to form the fine-grained energy efficient scheduling. 
{\color{black}However, the model accuracy can be further improved by using a 3-Dimensional computational fluid dynamics model rather than 2-Dimensional one.} 

\subsection{LOAD}
LOAD \cite{ranjbari2018learning} is an algorithm based on learning automata for energy and SLA efficient consolidation of VMs in cloud data centers. The proposed algorithm considers the demanded resource of users to predict overloaded hosts. By preventing overloaded hosts and shutting down idle hosts, the proposed algorithm aims to save energy consumption of data centers. The learning automata-based overload detection enhances the VM consolidation by predicting CPU usage of hosts upon resource usage history. Each VM is equipped with one automata including 3 actions, increasing CPU utilization, reducing CPU utilization and no changing of CPU utilization. At the beginning, the 3 actions are with equal probability. After the beginning stage, the reward and penalty of learning automata will be updated based on the environment. The automata selects one of the three actions in each iteration based on the probability. And in the next iteration, if the automata's decision is right, the action will be rewarded, otherwise, the action will be penalized. The learning automata is applied to estimate usage of VMs on the host. If the estimation shows that the host may be overloaded, the VMs will be migrated and other VMs will be prevented to be migrated to the current host. The migrated destination is based on the Best Fit Decreasing (BFD) algorithm \cite{Yue1991}. The simulation results show that leveraging learning-based prediction can reduce energy consumption of data centers. 

LOAD is a typical algorithm applying learning techniques to optimize VM consolidation. This work advances the existing work by considering the dynamic prediction for resource usage. But the limitation is that this work predicts the overloaded situations and does not handle the under-utilized situations.

\subsection{ACS}
ACS \cite{ACW2015} is an online optimization meta-heuristic algorithm based on ant colony optimization and VM consolidation to achieve the near-optimal solution. Its objective is to balance the energy consumption, the number of VM migrations and QoS concerning performance. In this approach, the authors formulate the energy efficient VM consolidation as a multi-objective optimization problem to optimize multiple metrics simultaneously. To leverage ACO, the necessary entities, including pheromone updating rules and probabilistic decision rules are defined. If one solution has more pheromone trails, the VM has a larger probability to be placed on the host. ACS also has updating rules for local and global pheromone, which are applied in each iteration. In iterations, the local pheromone is updated by ant when they perform a movement. After all ant construct their solutions locally, the global pheromone update is conducted as the migration process, and only the dominated place will be kept. The process is repeated until it reaches the maximum iteration number.

 ACS represents a set of meta-heuristic algorithms proposed to balance multiple objectives. The results based on simulations show that the proposed algorithm can reduce energy consumption and VM migrations while guaranteeing QoS. The performance can be further improved by investigating the impact of VM migration on network. 

In summary, the investigated algorithms all follow the two-phase energy efficient scheduling process as shown in Fig. \ref{fig:vmconsolidation}. The investigated algorithms apply different techniques to optimize the placement of VMs. Most of the investigated algorithms focus on the optimization of VM consolidation phase except EcoCloud spends more effort on the initial placement based on probabilistic approach. In the algorithms focusing on the VM consolidation optimization, MBFD applies heuristic algorithms based on bin-packing modelling, while EcoCloud utilizes a probabilistic approach to find the host with the highest probability to accept migrated VM. GRANITE considers energy and performance together by modelling cooling energy consumption and VM performance degradation. ACS applies a meta-heuristic algorithm to find the optimized consolidation solutions. 


\section{Architecture and Modelling}

\begin{figure}[!h]
	\centering	\includegraphics[width=0.75\linewidth]{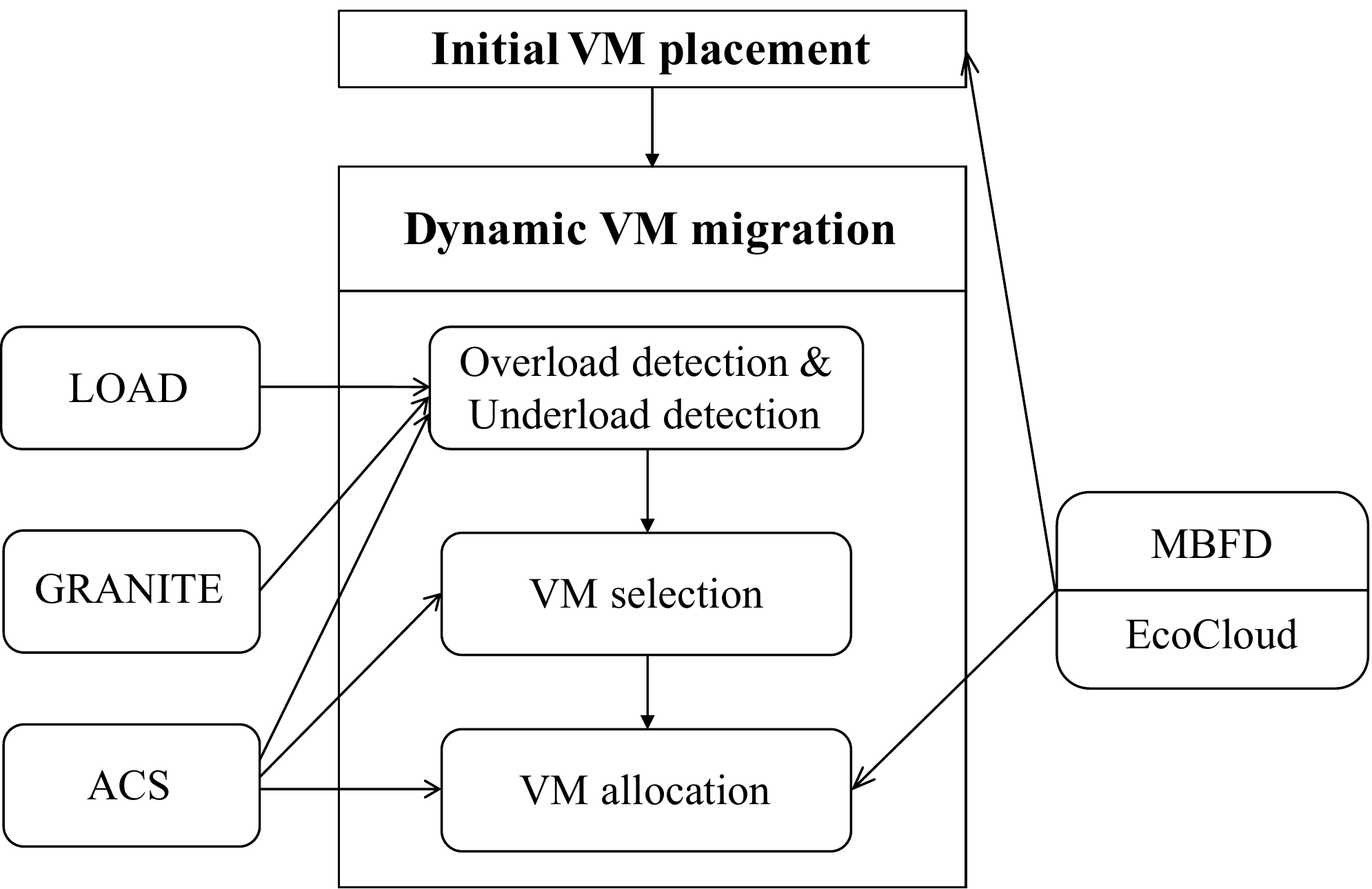}
	\caption[VarPerOptCom]{The focused scheduling phase comparison of investigated algorithms}
	\label{fig:focusedSchedulingPhase}
\end{figure}

This section discusses the investigated algorithms from the architecture and modelling perspectives, and the algorithm complexity is also discussed. 

\subsection{MBFD} 

\textbf{Architecture:} The four main components of the green cloud architecture are: broker, green service allocator, Virtual Machines (VMs) and Physical Machines (PMs). Broker enables the user interaction module to submit the workloads and their QoS requirements from any geographical distribution allocation. PM based hardware infrastructure creates virtualized resources (VMs). VMs are consolidated to fulfill the demand of workloads dynamically using DVFS. Green service allocator incorporates the energy manager and VM manager to allocate the virtual resources to user workloads based on their requirements for their execution at runtime. 

\textbf{Model:} Equation (\ref{eq:MBFP}) defines the power model for this research work.
\begin{equation}
\label{eq:MBFP}
P(u)=k \times P_{max} + (1-k)\times P_{max}\times u
\end{equation}
where $P_{max}$ is the maximum consumption of power while fully utilization of server; $k$ is the small amount of power used by idle server; and $u$ is the CPU utilization. The value of energy consumption $E$ of a PM is defined in Equation (\ref{eq:MBFD1}). $u(t)$ is CPU utilization, which is a function of time. 
\begin{equation}
\label{eq:MBFD1}
E=\int_{t2}^{t1}P(u(t))dt
\end{equation}
where $t1$ is start time of Task $T$ and $t2$ is end time of Task $T$.


\subsection{EcoCloud}


\textbf{Architecture:} The architecture of EcoCloud contains two main probabilistic procedures: 1) VM assignment and VM migration. Based on the availability of CPU and RAM on different servers of cloud data centers, VM allocation is performed. EcoCloud allocates the VM to the newly submitted application and sends an invitation to all the servers participated to find the best one to place this VM. 

\textbf{Model:} Equation (\ref{eq:ecocloud}) defines the linear model used in this research for power consumption and it is expressed as:
\begin{equation}
\label{eq:ecocloud}
P(u) = P_{idle} + (P_{max}-P_{idle})\times u
\end{equation}
where $P_{max}$ is the maximum consumption of power while fully utilization of server; $P_{idle}$ is the small amount of power used by idle server; and $u$ is the CPU utilization.


\subsection{GRANITE}

\textbf{Architecture:} It comprises three sub-components including workload manager, scheduling manager and cooling manager. Workload manager manages the workloads submitted by users and process for scheduling based on their requirements. Scheduling manager schedules the resources for the execution of workloads while maximizing the performance of data centers and minimizes the consumption of energy. Cooling manager maintains the temperature of data centers and saves cooling energy by performing VM placement and dynamic migration in an efficient manner. 

\textbf{Model:} Equation (\ref{eq:granite}) defines a linear power model to find the energy consumption of data centers, which is a combination of computing energy and cooling energy. 
\begin{equation}
\label{eq:granite}
Energy = Energy_{computing} + Energy_{cooling}
\end{equation}


\subsection{LOAD}

\textbf{Architecture:} The system architecture contains four sub-components such as user portal (to submit workloads), global manager (which is an intermediate between user portal and local manager), local manager (which manages the PMs and VMs, and it is controlled by a single global manager) and VM manager (which manages the virtual resources). 

\textbf{Model:} Equation (\ref{eq:MBFP}) is used to calculate the energy consumption for this research work, which represents the linear model of power and CPU utilization.


\subsection{ACS}


\textbf{Architecture:} In the architecture of ACS, two types of agents are included: local and global agents. The local agent is deployed in a host to solve the host status detection sub-problem by monitoring the host resource utilization. The global agent is responsible for supervising and optimizing the VM placement by taking advantage of the ACO-based algorithm. 

\textbf{Model:} ACS utilizes the linear power model as MBFD in Equation \ref{eq:MBFD1}. \\

		We note that the investigated algorithms have different focuses on the phases of VM consolidation, and we show the focuses in Fig.  \ref{fig:focusedSchedulingPhase}. As introduced, the VM consolidation process can be mainly divided into two phases: the initial VM placement and dynamic VM migration. And for the VM migration, the overloads and under-utilized detection, VM selection and VM allocation are included. 
		MBFD and EcoCloud spend more effort on the VM placement phase, including the initial VM placement and VM allocation in Dynamic VM migration. On the other hand, LOAD focuses on overloads detection by predicting the utilization via learning automaton to optimize overload detection. GRANITE takes the CPU temperature into consideration and optimizes the overload detection policy. ACS improves all the phases in dynamic VM consolidation. It utilizes LiRCUP\cite{farahnakian2013lircup} to forecast overloads of servers, and adopts Ant Colony System to find the near-optimal solutions for VM selection and allocation.

In summary, from the perspective of architecture, the investigated algorithms all adopt the layered architecture, and the layers can be mainly divided into three parts. The bottom layer is the resource provisioner, which provides physical or virtual resources. The middle layer is responsible for the energy efficient scheduling, which handles the VM management and provides the energy efficient scheduling algorithms. At the top layer, the users' request and optimization goals are configured. 

From the modelling perspective, all algorithms adopt the linear power model. As for power components, all algorithms include the CPU energy consumption, MBFD, EcoCloud and ACS consider the memory part, and GRANITE utilizes a more comprehensive model including storage, network and cooling. 

\textbf{Algorithm complexity analysis:} MBFD, GRANITE and LOAD are based on heuristic algorithms, and the complexity of them are all $M\times N$, where $N$ is the number of hosts and $M$ is the number of VMs. The complexity of EcoCloud equals to the number of hosts to accept VMs, which is $N$. Based on meta-heuristic with iterations, the complexity of ACS is $M\times N \times A\times I$, where $A$ is the number of ants that concurrently build their migration plans, and $I$ is the number of iterations.

 
\section{Metrics}
For the objective of energy efficiency, energy consumption is the major metric to be evaluated. However, the algorithms are also making trade-offs between the energy consumption and other metrics, such as SLA violations. In this section, we discuss the adopted metrics in our investigated energy efficient algorithms. Note that the investigated algorithms use some similar metrics while having some other additional metrics. Here we aim to cover the metrics applied in these algorithms and identify the differences between these metrics. Table \ref{tab:metrics} summaries the algorithms with their corresponding adopted metrics. 

\textit{Metrics for energy efficiency}

\textbf{Total energy consumption:} It is the total energy consumption consumed by physical machines in the data centers. It is derived from the power model in Equation (\ref{eq:MBFD1}).

\textbf{Number of active servers:} It represents how many servers are running as active during the observation time. The value should be minimized, and thus more idle serves can be switched into low-power mode.

\textit{Metrics for SLA}

\textbf{SLA violation percentage (SLAV) \cite{beloglazov2012energy}}: It is the percentage of SLA violations events relatively to the total number of the processed time frame. The SLA violation is identified when a given VM cannot get the amount of MIPS as requested. 

\textbf{VM migration times}: The number of migrations triggered by the algorithm during the VM scheduling process. 

\textbf{Average SLA violation}: It is the average CPU performance that has not been allocated to an application when requested, resulting in performance degradation.





\begin{table}[]
	\caption{Metrics adopted in compared algorithms}
	\label{tab:metrics}
	\resizebox{0.48\textwidth}{!}{%
		\begin{tabular}{|c|c|c|}
			\hline
			Metrics & Optimization Objective & Adopted Algorithms \\ \hline
			Total Energy Consumption & Minimization & All \\ \hline
			SLA violation percentage & Minimization & All \\ \hline
			VM migrations & Minimization & BMDP, LOAD, EcoCloud, ACS \\ \hline
			Average SLA violation & Minimization & BMDP, LOAD, ACS \\ \hline
			Number of active hosts & Minimization & EcoCloud, GRANITE\\ \hline
		\end{tabular}%
	}
\end{table}


In summary, we can notice that several metrics have been adopted for evaluations by more than one algorithm, including total energy, SLA violation percentage, VM migrations, average SLA violations and number of active hosts. To make our evaluations more comparable from the metrics perspective, we evaluate these metrics in the performance evaluations section.

\begin{table*}[]
	\centering
	\caption{Energy consumption with different CPU utilization {\color{black}in Watts}}
	\label{tab:power_model}

	\begin{tabular}{|l|ccccccccccc|}
		\hline
		\textbf{Server} & 0\% & 10\% & 20\% & 30\% & 40\% & 50\% & 60\% & 70\% & 80\% & 90\% & 100\% \\
		\hline
		\textbf{HP ProLiant G4} & 86 & 89.4 & 92.6 & 96 & 99.5 & 102 & 106 & 108 & 112 & 114 & 117 \\
		\textbf{HP ProLiant G5} & 93.7 & 97 & 101 & 105 & 110 & 116 & 121 & 125 & 129 & 133 & 135 \\ \hline
	\end{tabular}

\end{table*}

\section{Performance Evaluations}

\begin{figure*}[t]
	\centering
	\begin{subfigure}{0.24\linewidth}
		\centering
		\includegraphics[width=0.99\linewidth]{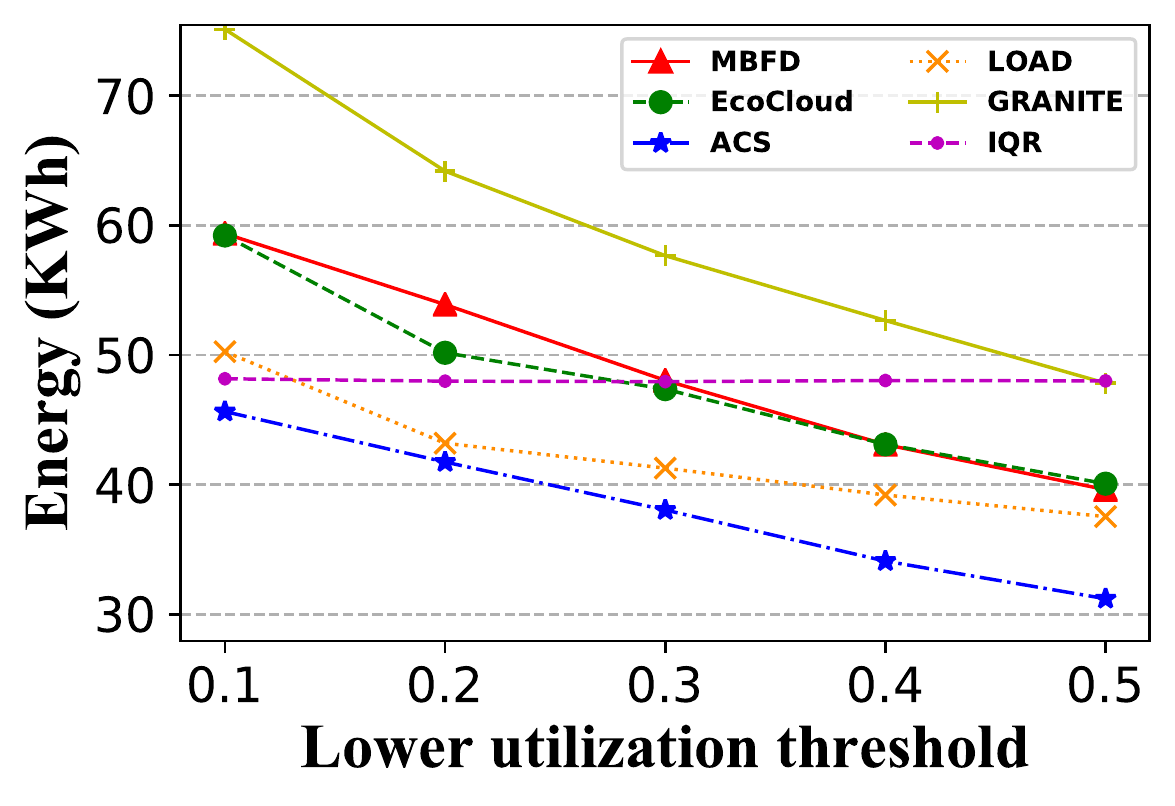}
		\caption{Energy consumption}
		\label{fig:energyConsumptionVaryingThreshold}
	\end{subfigure}
	\begin{subfigure}{0.24\linewidth}
		\centering
		\includegraphics[width=0.99\linewidth]{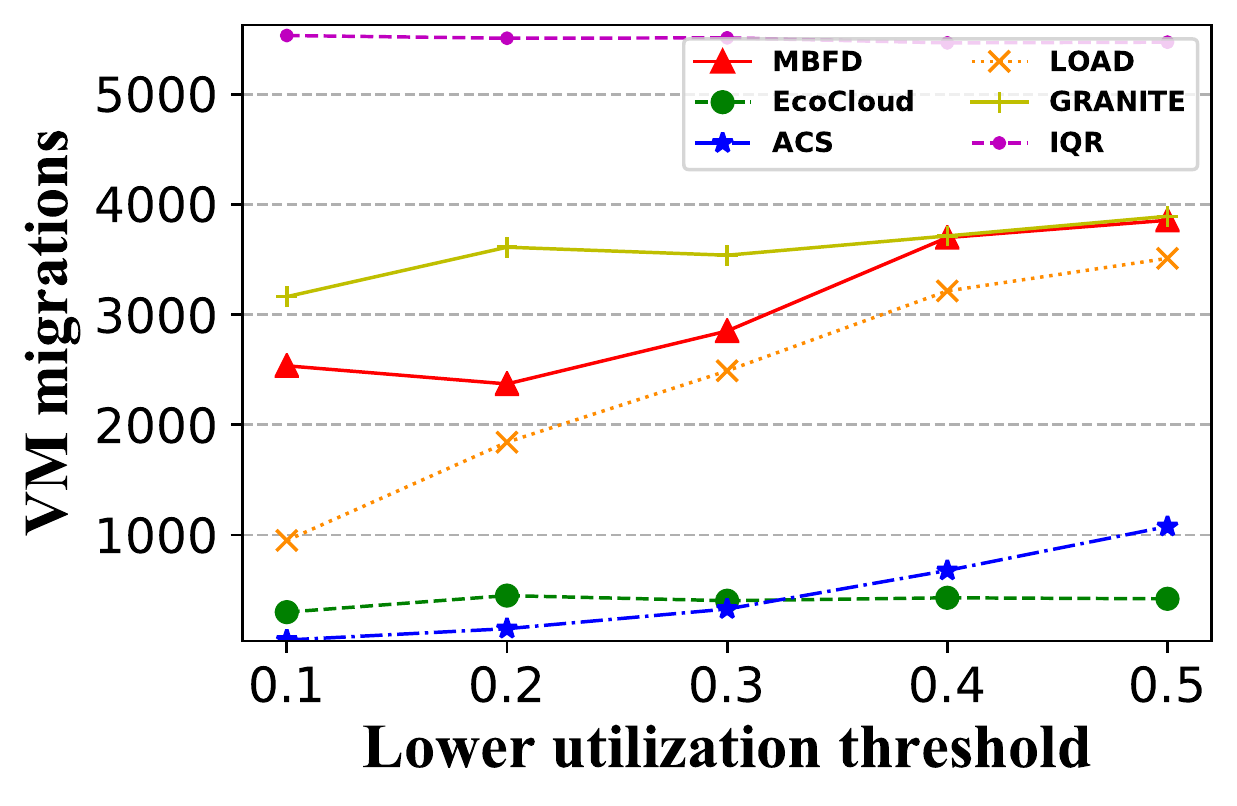}
		\caption{VM migrations}
		\label{fig:VMmigrationVaryingThreshold}
	\end{subfigure}
	\begin{subfigure}{0.24\linewidth}
		\includegraphics[width=0.99\linewidth]{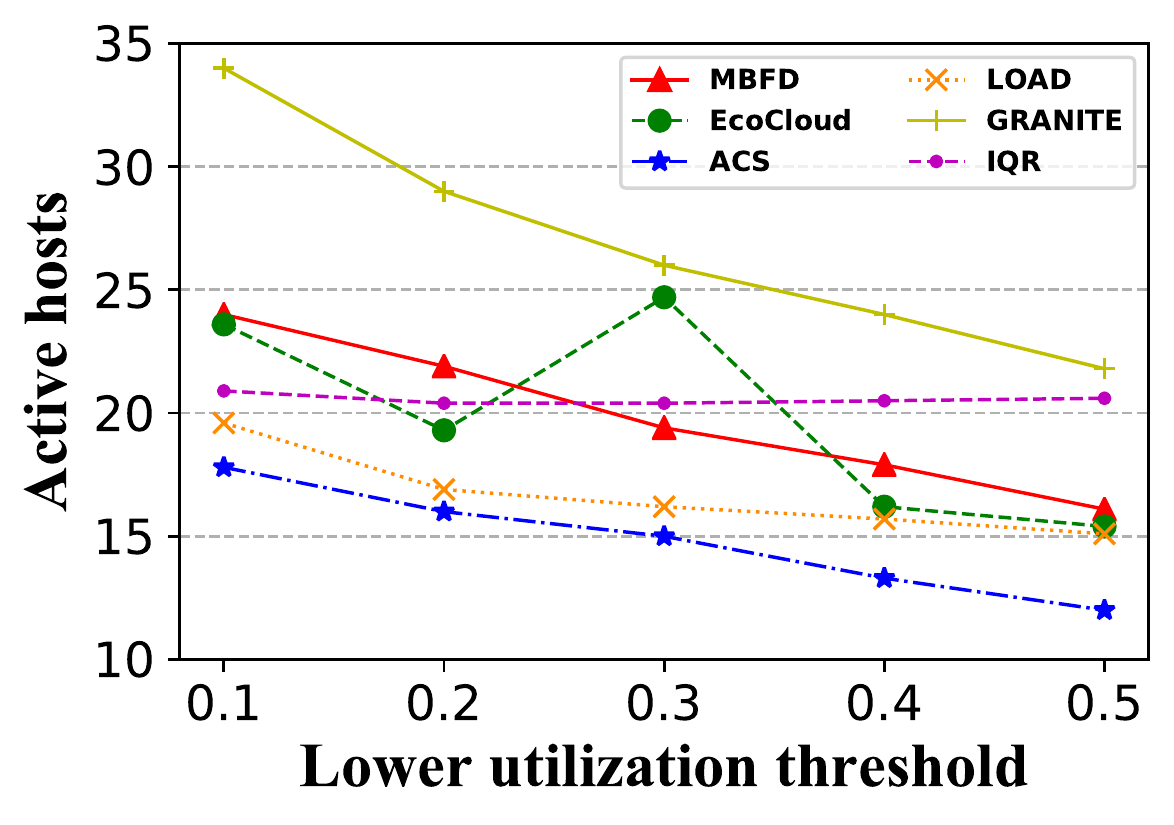}
		\caption{Number of active hosts}
		\label{fig:numberActiveHostVaryingThreshold}
	\end{subfigure}
	\begin{subfigure}{0.24\linewidth}
		\includegraphics[width=0.99\linewidth]{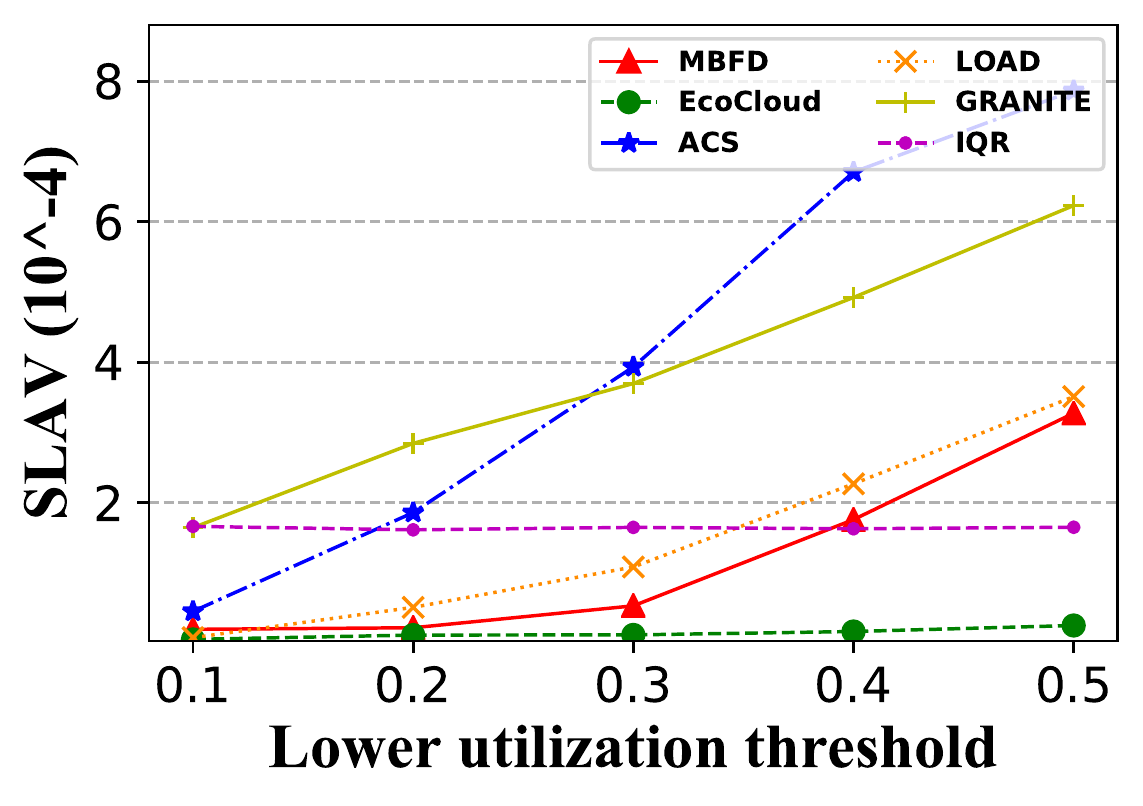}
		\caption{SLAV}
		\label{fig:SLAVVaryingThreshold}
	\end{subfigure}
	
	\caption[VarPerOptCom]{Performance comparison of algorithms under {\color{black}Synthetic} workloads (ratio of hosts and VMs number is \color{black}{50:50})}
	\label{fig:random1.0}
	
\end{figure*}

\begin{figure*}[t]
	\centering
	\begin{subfigure}{0.24\linewidth}
		\centering
		\includegraphics[width=0.99\linewidth]{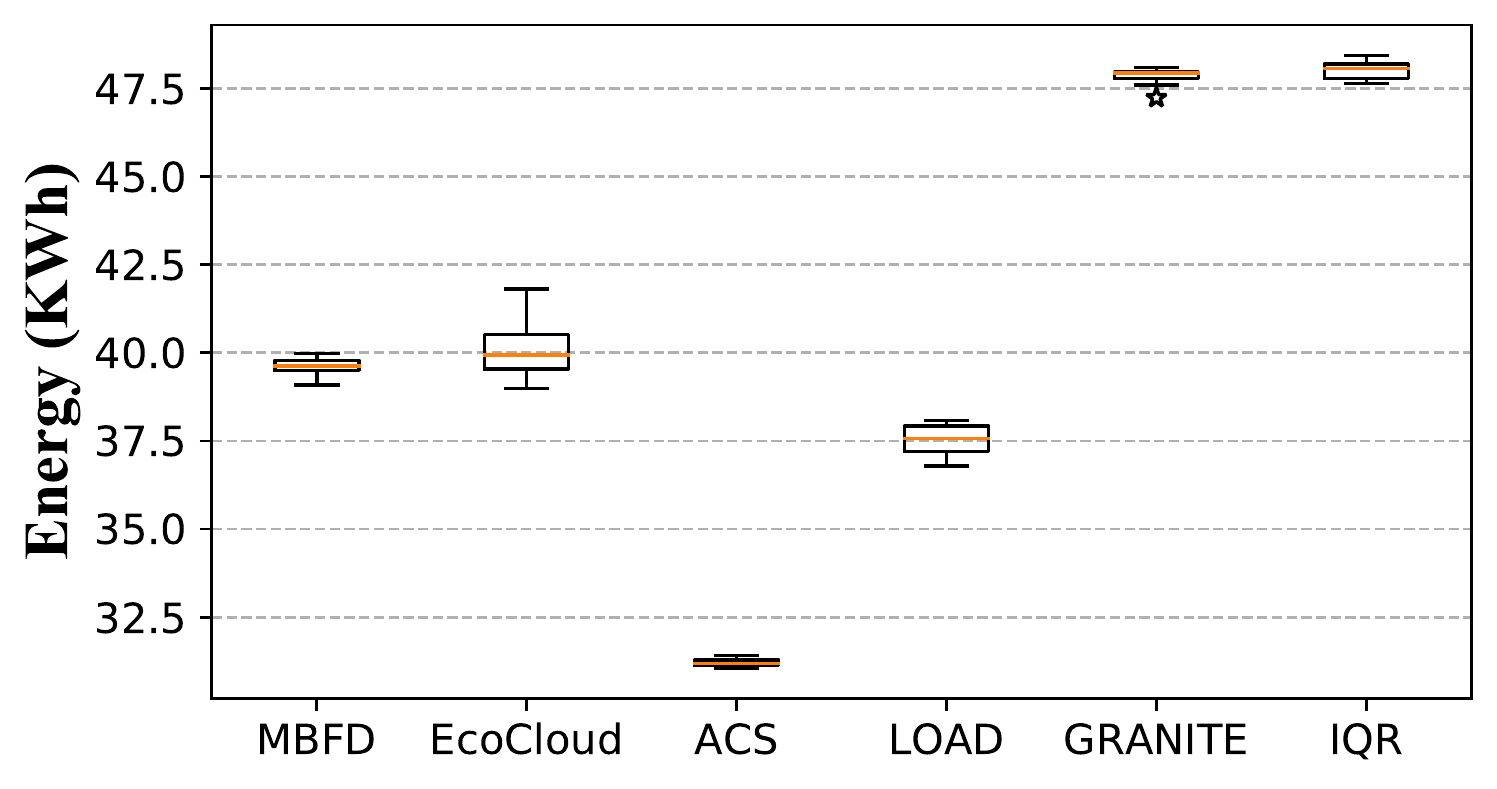}
		\caption{Energy consumption}
		\label{fig:boxplotEnergyConsumption}
	\end{subfigure}
	\begin{subfigure}{0.24\linewidth}
		\centering
		\includegraphics[width=0.99\linewidth]{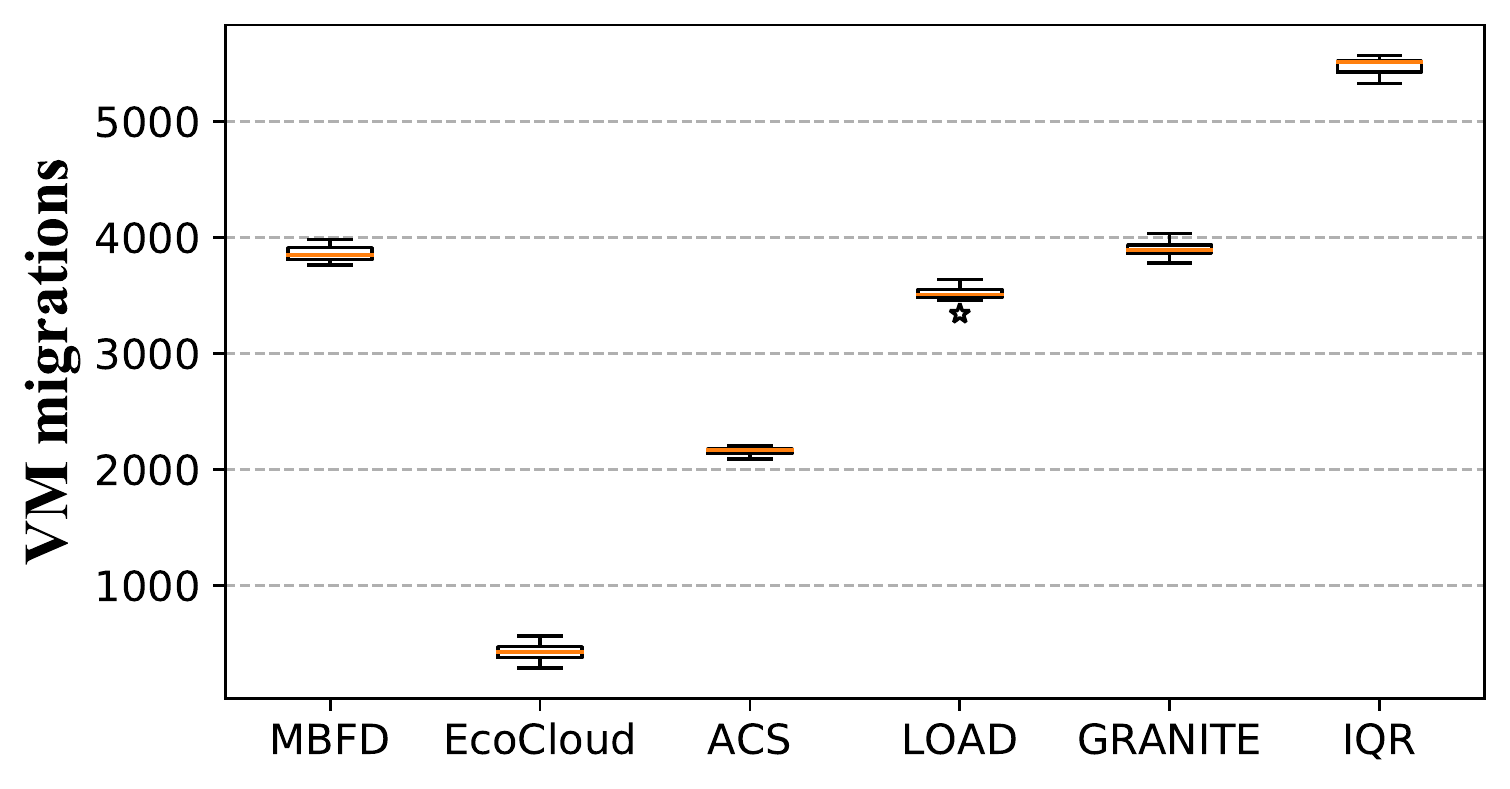}
		\caption{VM migrations}
		\label{fig:boxplotVMmigration}
	\end{subfigure}
	\begin{subfigure}{0.24\linewidth}
		\includegraphics[width=0.99\linewidth]{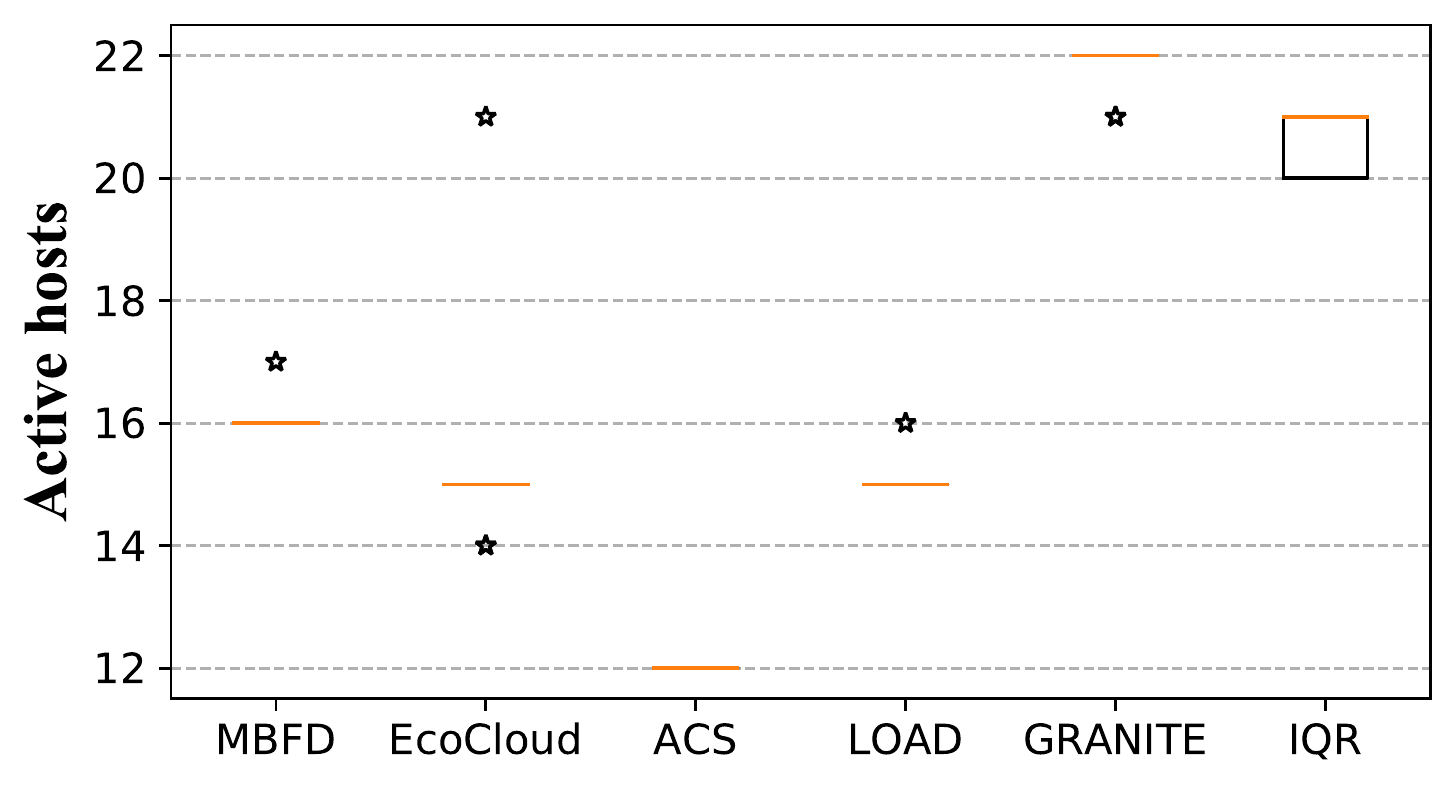}
		\caption{Number of active hosts}
		\label{fig:boxplotActiveHost}
	\end{subfigure}
	\begin{subfigure}{0.24\linewidth}
		\includegraphics[width=0.99\linewidth]{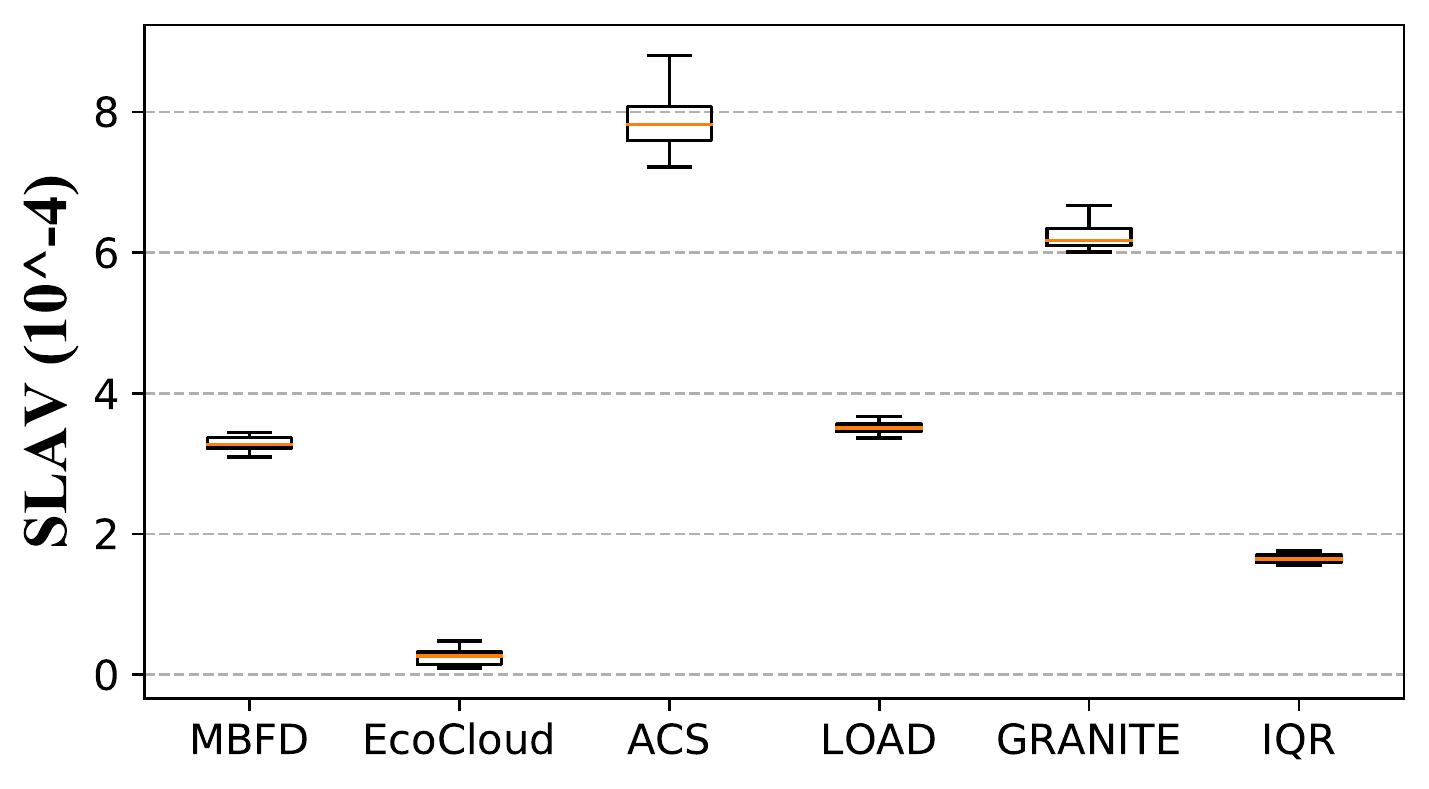}
		\caption{SLAV}
		\label{fig:boxplotSLAV}
	\end{subfigure}	
	\caption[VarPerOptCom]{Performance comparison of algorithms under {\color{black}Synthetic} workloads with lower utilization threshold as 0.5 (ratio of hosts and VMs number is \color{black}{50:50})}
	\label{fig:boxplotRandom1.0}
	
\end{figure*}

\begin{figure*}[t]
	\centering
	\begin{subfigure}{0.24\linewidth}
		\centering
		\includegraphics[width=0.99\linewidth]{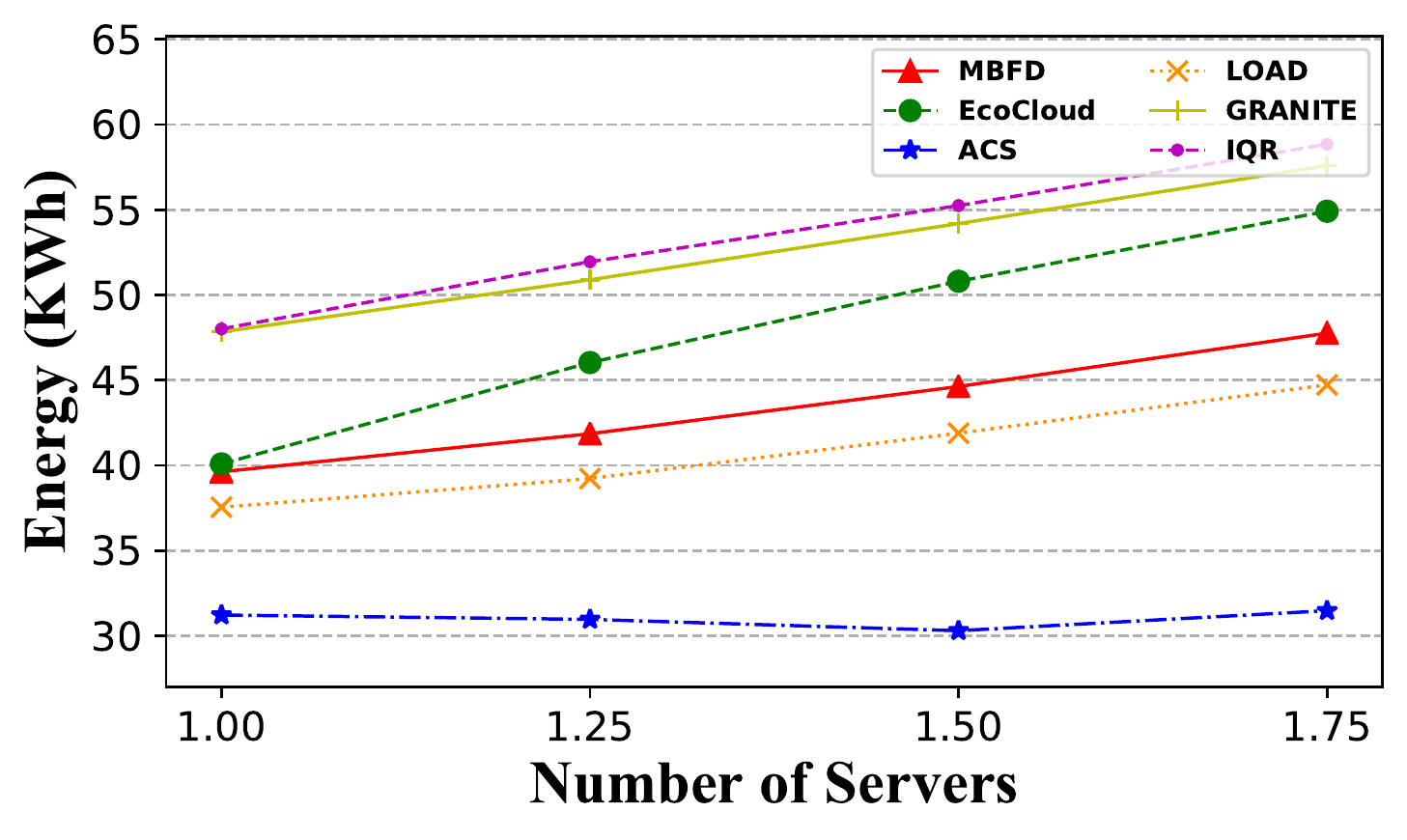}
		\caption{Energy consumption}
		\label{fig:energyConsumptionVaryingThresholdRandomRatio}
	\end{subfigure}
	\begin{subfigure}{0.24\linewidth}
		\centering
		\includegraphics[width=0.99\linewidth]{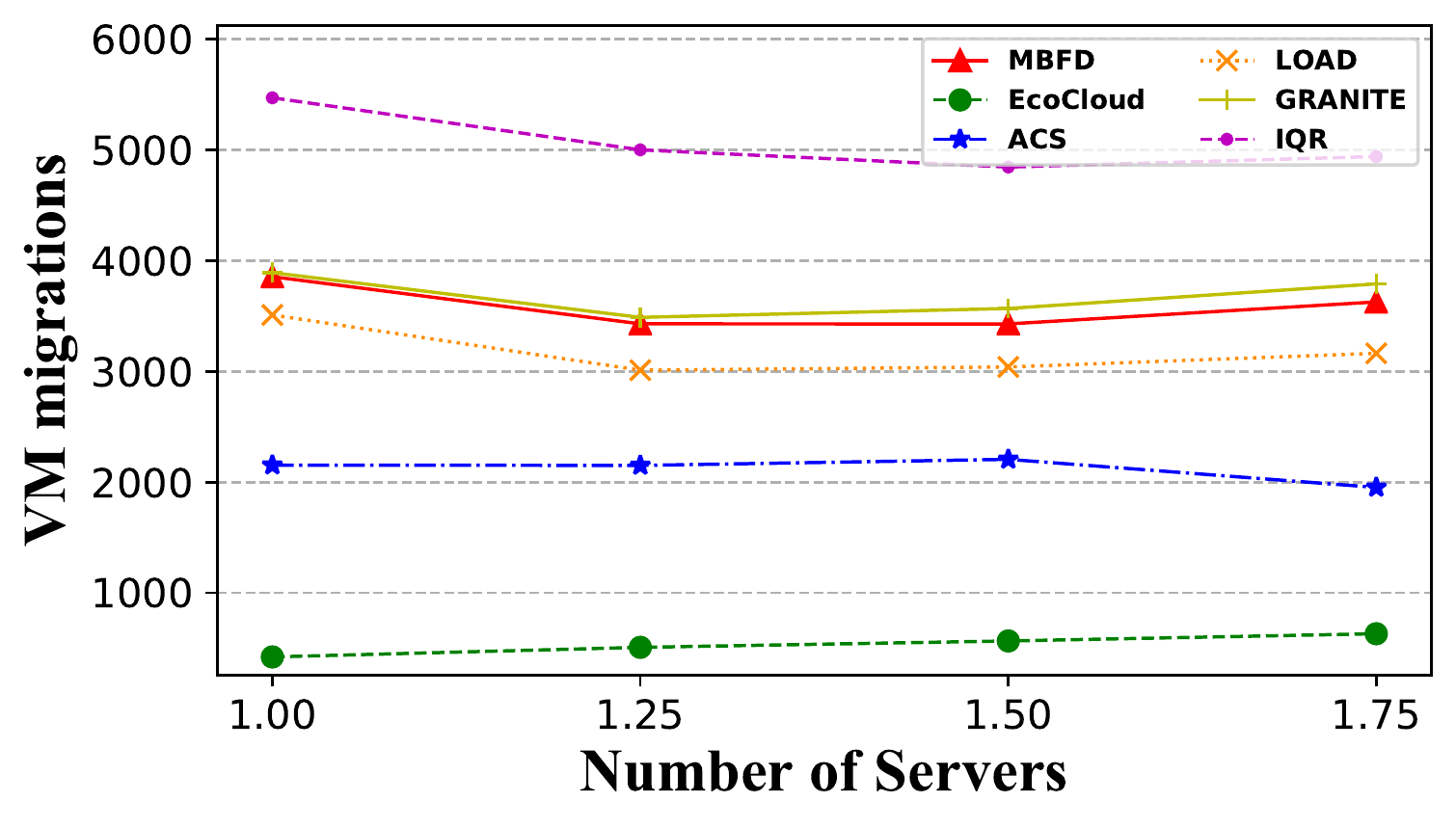}
		\caption{VM migrations}
		\label{fig:VMmigrationVaryingThresholdRandomRatio}
	\end{subfigure}
	\begin{subfigure}{0.24\linewidth}
		\includegraphics[width=0.99\linewidth]{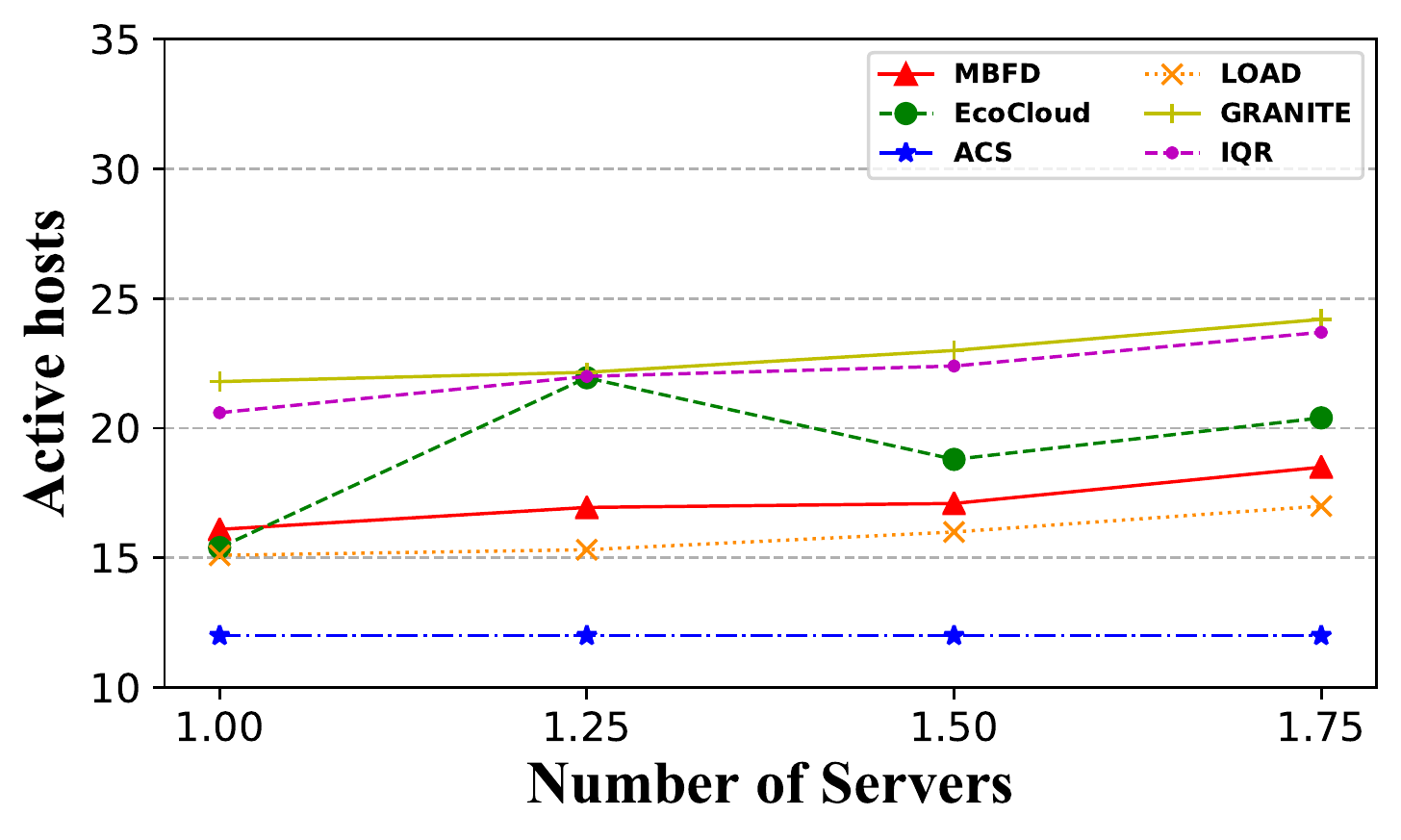}
		\caption{Number of active hosts}
		\label{fig:numberActiveHostVaryingThresholdRandomRatio}
	\end{subfigure}
	\begin{subfigure}{0.24\linewidth}
		\includegraphics[width=0.99\linewidth]{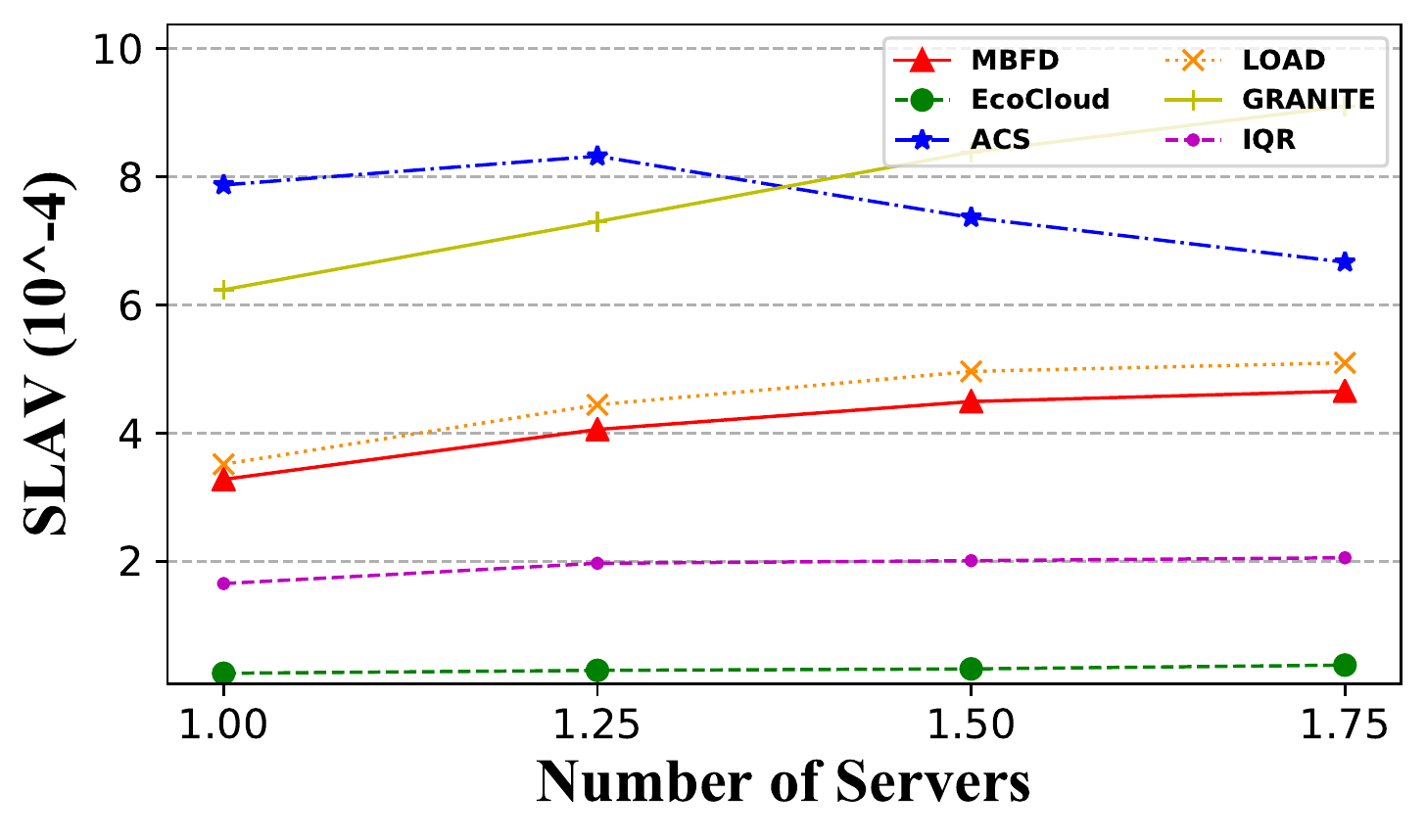}
		\caption{SLAV}
		\label{fig:SLAVVaryingThresholdRandomRatio}
	\end{subfigure}
	
	\caption[VarPerOptCom]{Performance comparison of algorithms under {\color{black}Synthetic} workloads ({\color{black}{setting hosts number = 50 and}} varying ratios of hosts and VMs number are 1:1, 1:1.25, 1:1.5, and 1:1.75)}
	\label{fig:randomRatio}
	
\end{figure*}

\begin{figure*}[t]
	\centering
	\begin{subfigure}{0.24\linewidth}
		\centering
		\includegraphics[width=0.99\linewidth]{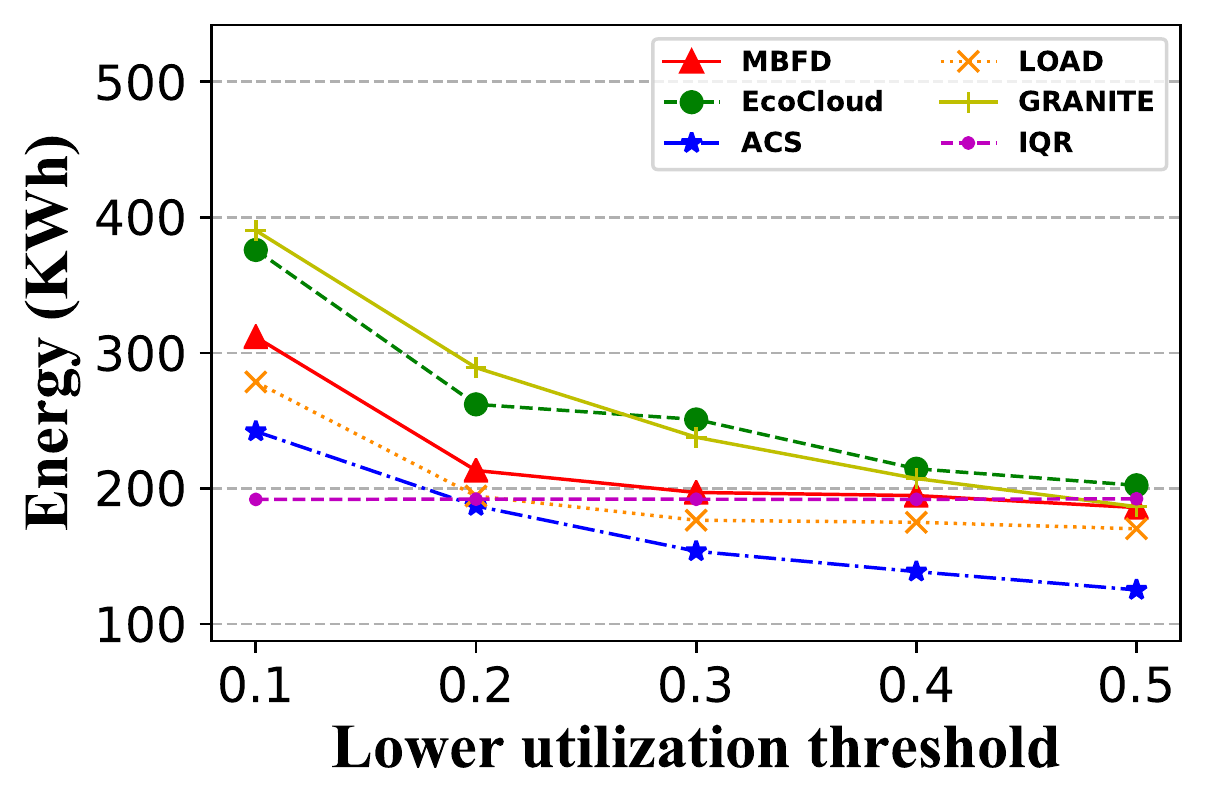}
		\caption{Energy consumption}
		\label{fig:energyConsumptionVaryingThresholdPlanet}
	\end{subfigure}
	\begin{subfigure}{0.24\linewidth}
		\centering
		\includegraphics[width=0.99\linewidth]{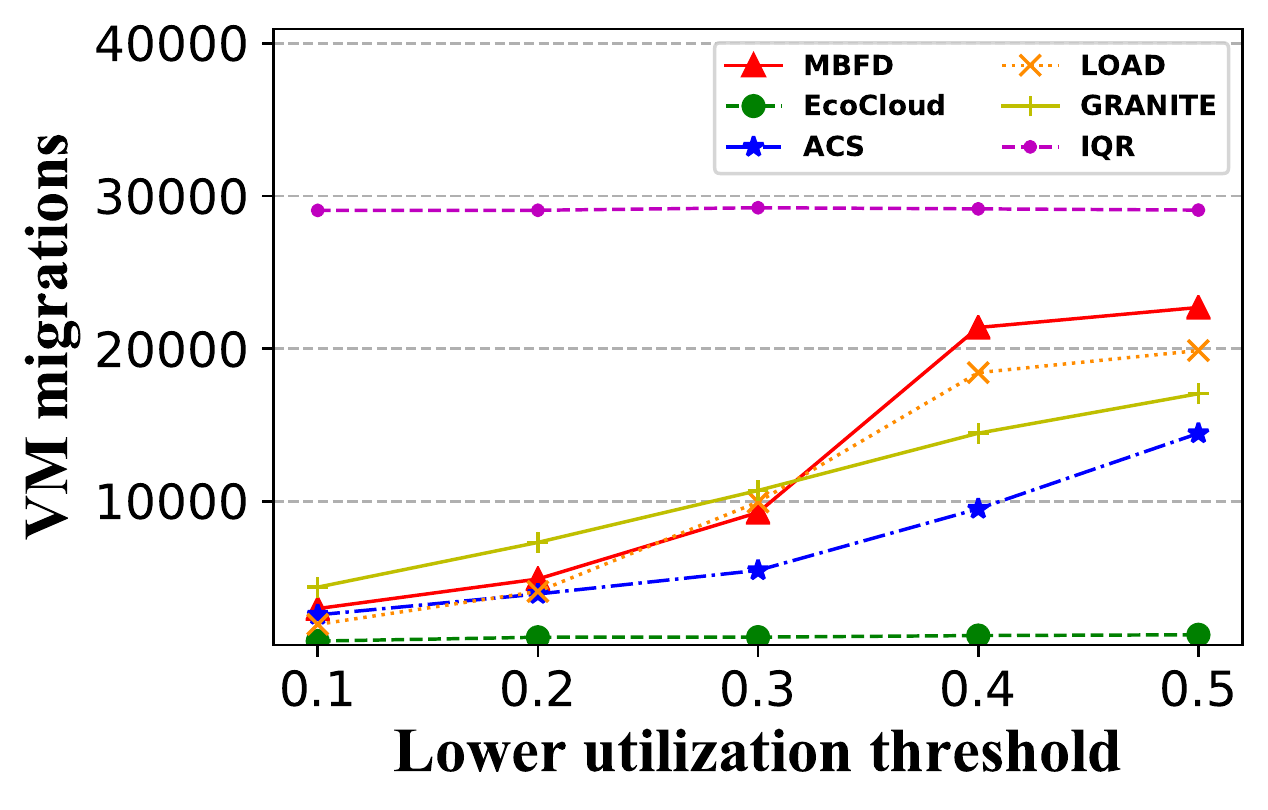}
		\caption{VM migrations}
		\label{fig:VMmigrationVaryingThresholdPlanet}
	\end{subfigure}
	\begin{subfigure}{0.24\linewidth}
		\includegraphics[width=0.99\linewidth]{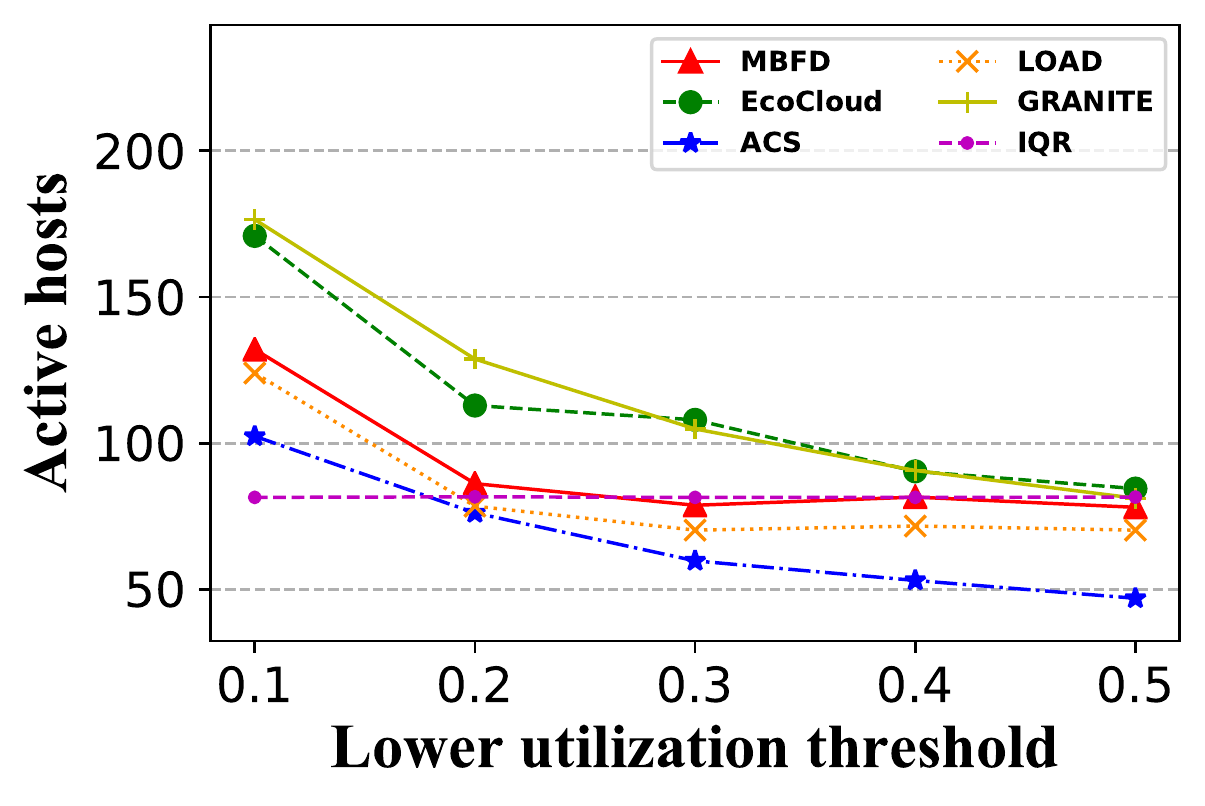}
		\caption{Number of active hosts}
		\label{fig:numberActiveHostVaryingThresholdPlanet}
	\end{subfigure}
	\begin{subfigure}{0.24\linewidth}
		\includegraphics[width=0.99\linewidth]{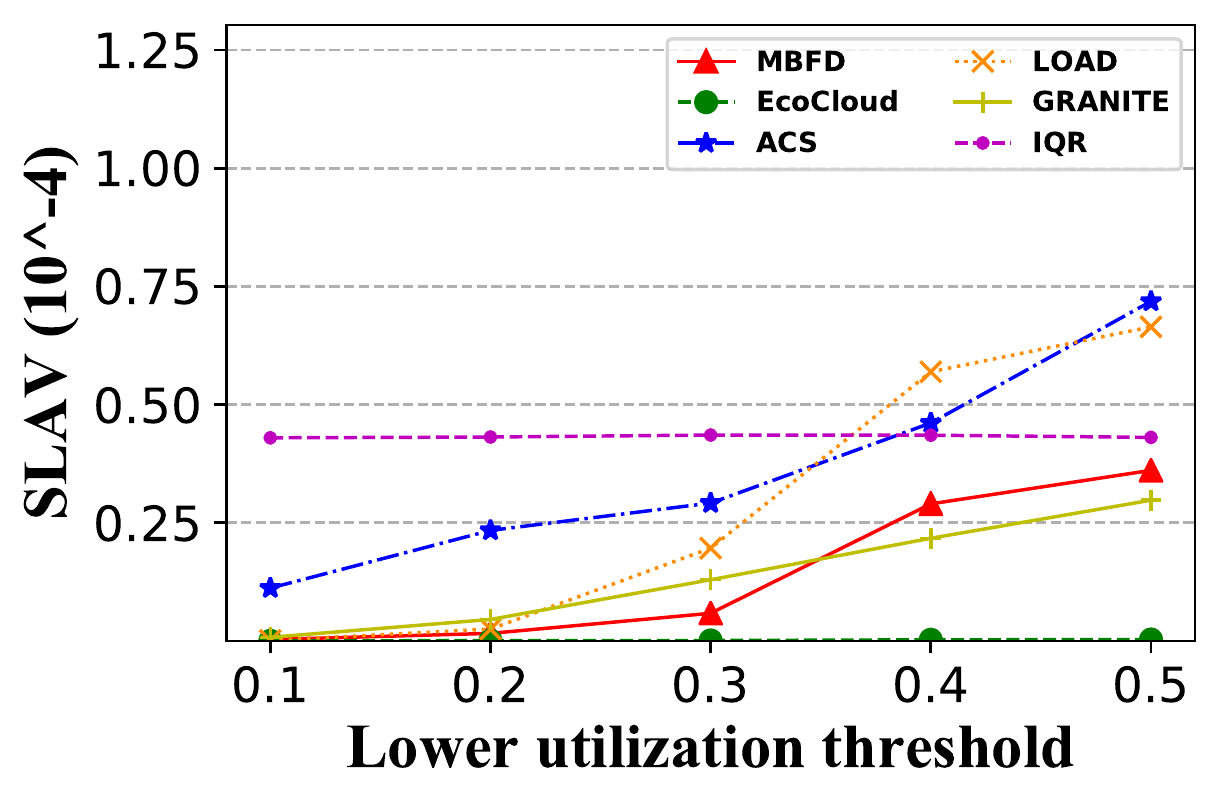}
		\caption{SLAV}
		\label{fig:SLAVVaryingThresholdPlanet}
	\end{subfigure}
	
	\caption[VarPerOptCom]{Performance comparison of algorithms under PlanetLab workloads}
	\label{fig:PerformanceVaryingThresholdPlanetlab}
	
\end{figure*}

\begin{figure*}[t]
	\centering
	\begin{subfigure}{0.24\linewidth}
		\centering
		\includegraphics[width=0.99\linewidth]{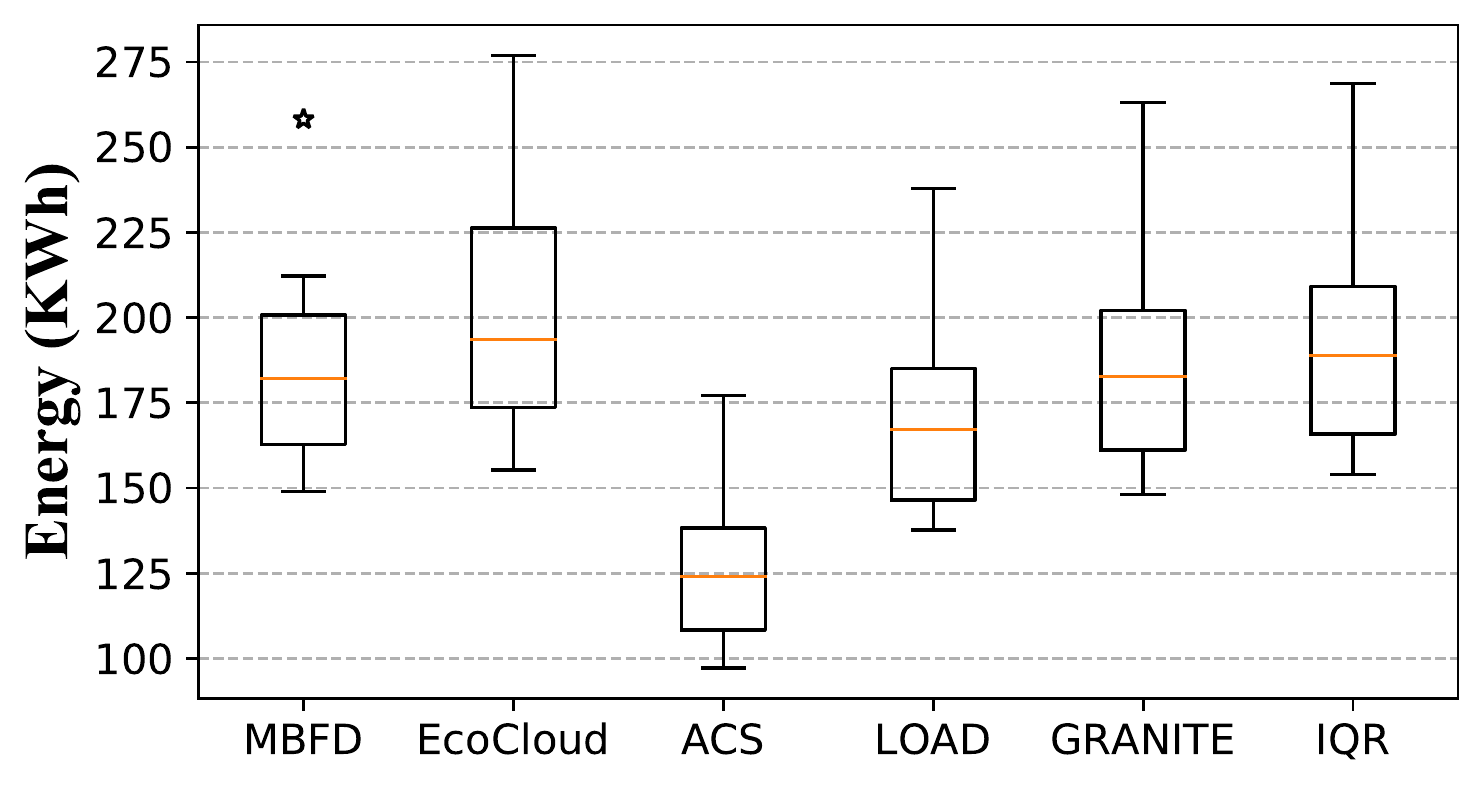}
		\caption{Energy consumption}
		\label{fig:boxplotEnergyConsumptionPlanet}
	\end{subfigure}
	\begin{subfigure}{0.24\linewidth}
		\centering
		\includegraphics[width=0.99\linewidth]{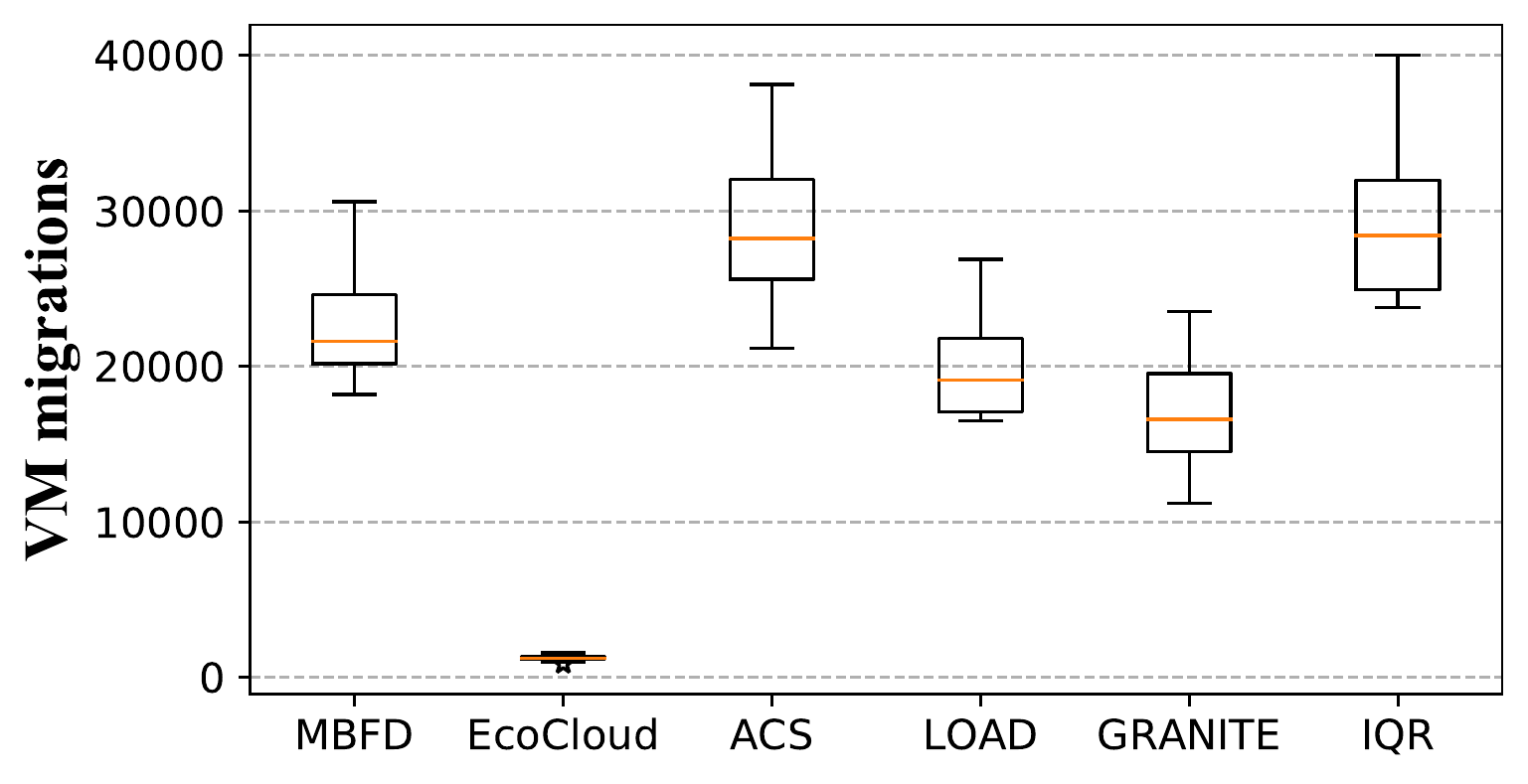}
		\caption{VM migrations}
		\label{fig:boxplotVMmigrationPlanet}
	\end{subfigure}
	\begin{subfigure}{0.24\linewidth}
		\includegraphics[width=0.99\linewidth]{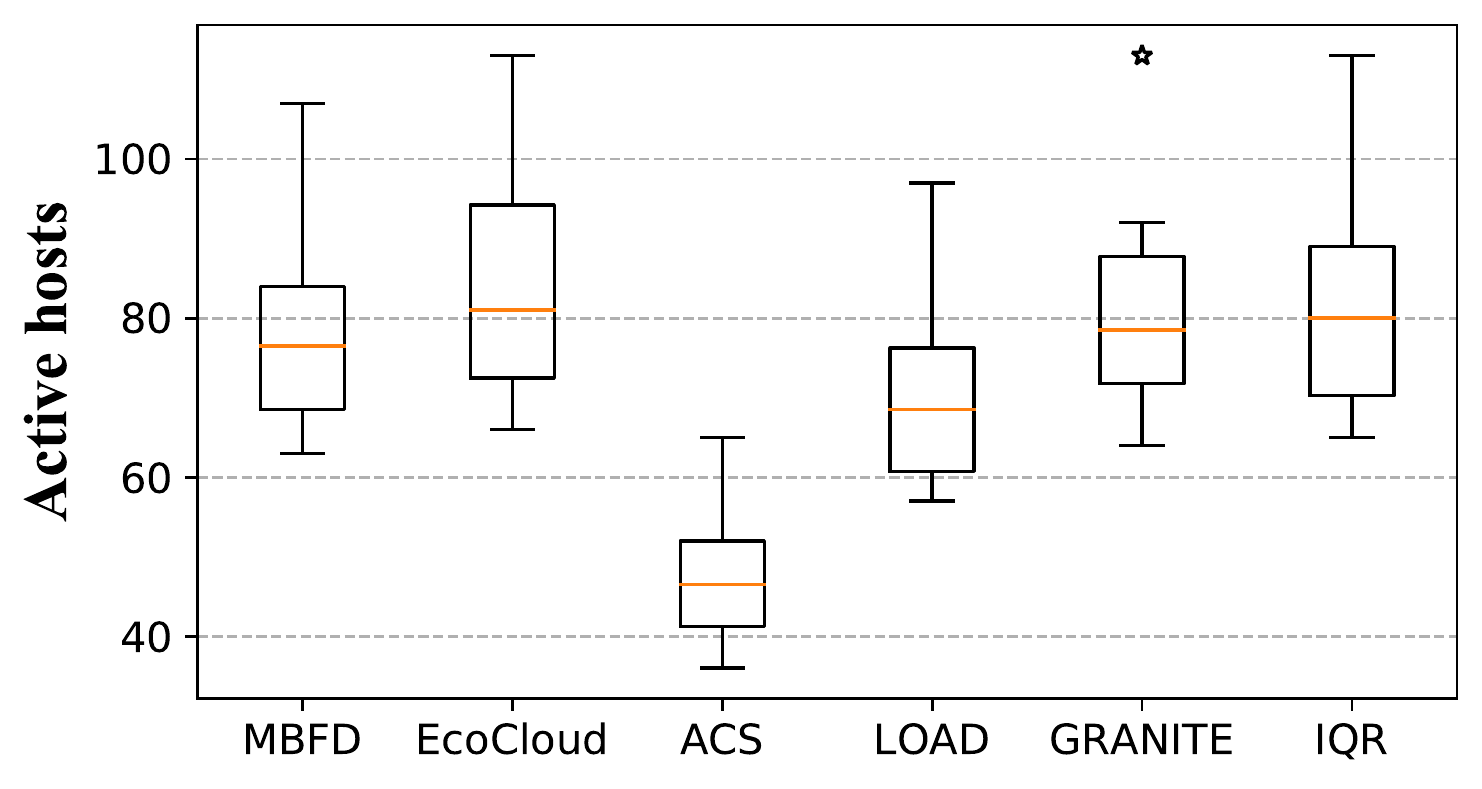}
		\caption{Number of active hosts}
		\label{fig:boxplotActiveHostPlanet}
	\end{subfigure}
	\begin{subfigure}{0.24\linewidth}
		\includegraphics[width=0.99\linewidth]{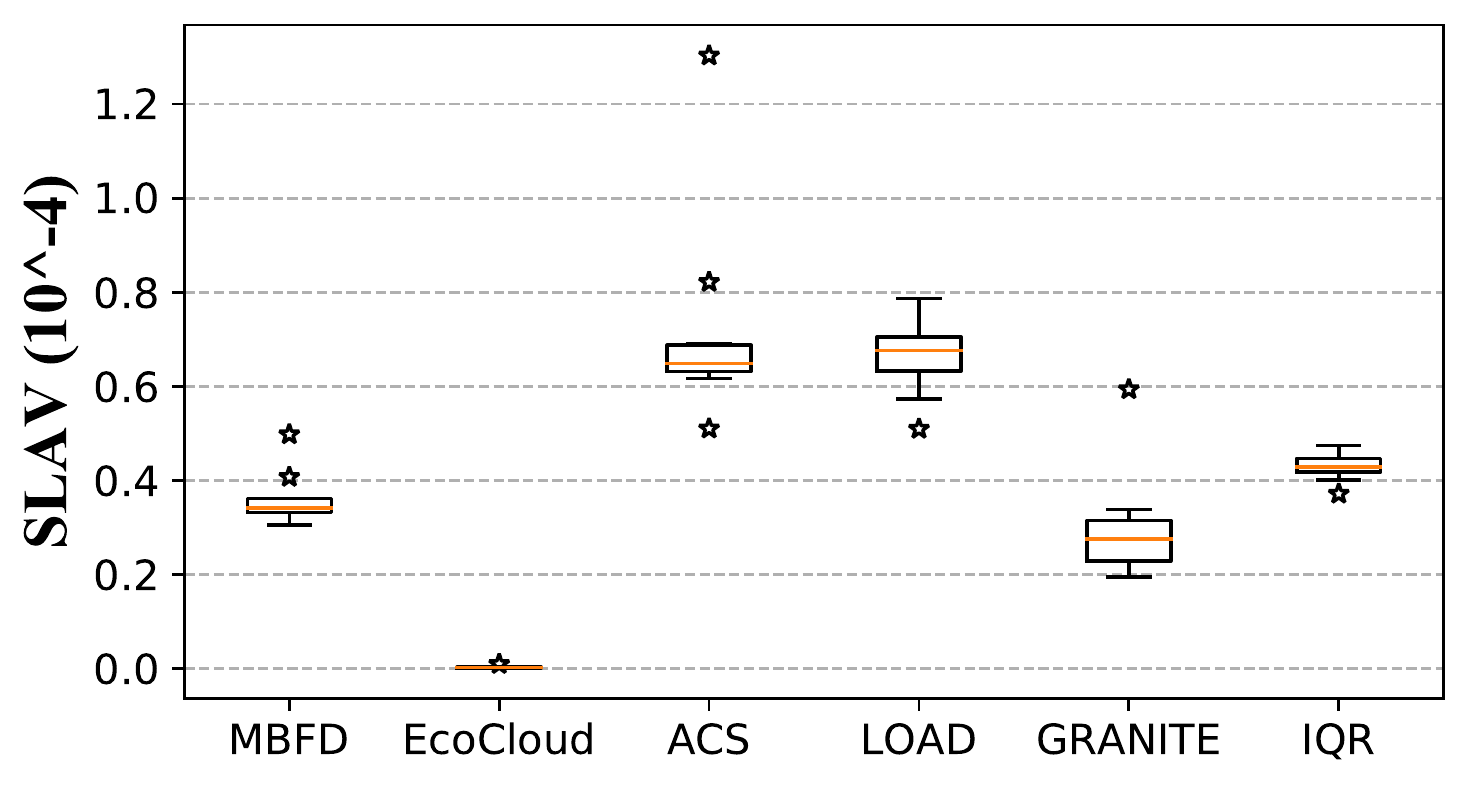}
		\caption{SLAV}
		\label{fig:boxplotSLAVPlanet}
	\end{subfigure}	
	\caption[VarPerOptCom]{Performance comparison of algorithms under PlanetLab workloads with lower utilization threshold as 0.5}
	\label{fig:PerformanceVaryingThreshold_boxplot_Planetlab}
	
\end{figure*}


\begin{figure*}[t]
	\centering
	\begin{subfigure}{0.24\linewidth}
		\centering
		\includegraphics[width=0.99\linewidth]{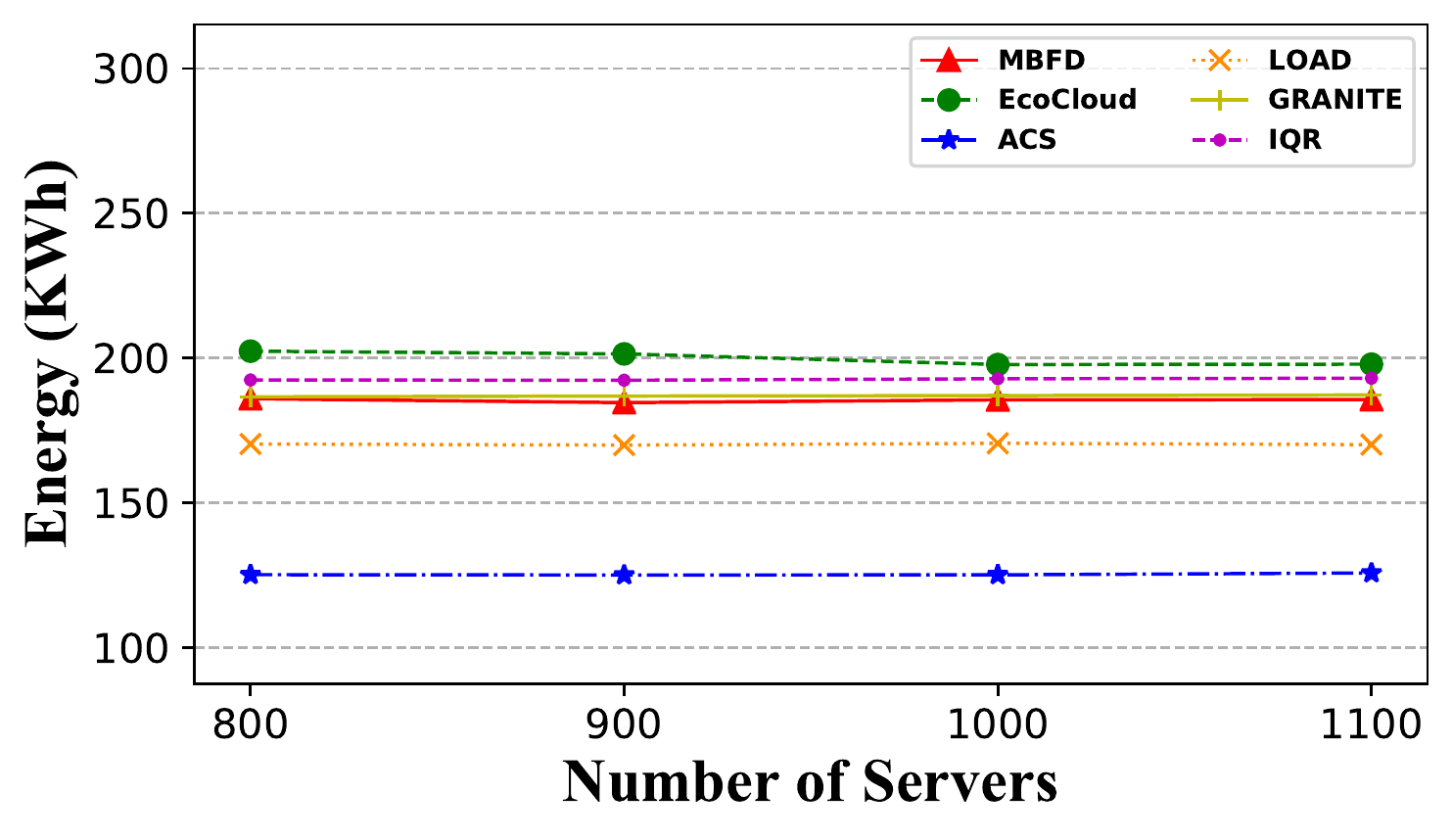}
		\caption{Energy consumption}
		\label{fig:energyConsumptionVaryingThresholdPlanetRatio}
	\end{subfigure}
	\begin{subfigure}{0.24\linewidth}
		\centering
		\includegraphics[width=0.99\linewidth]{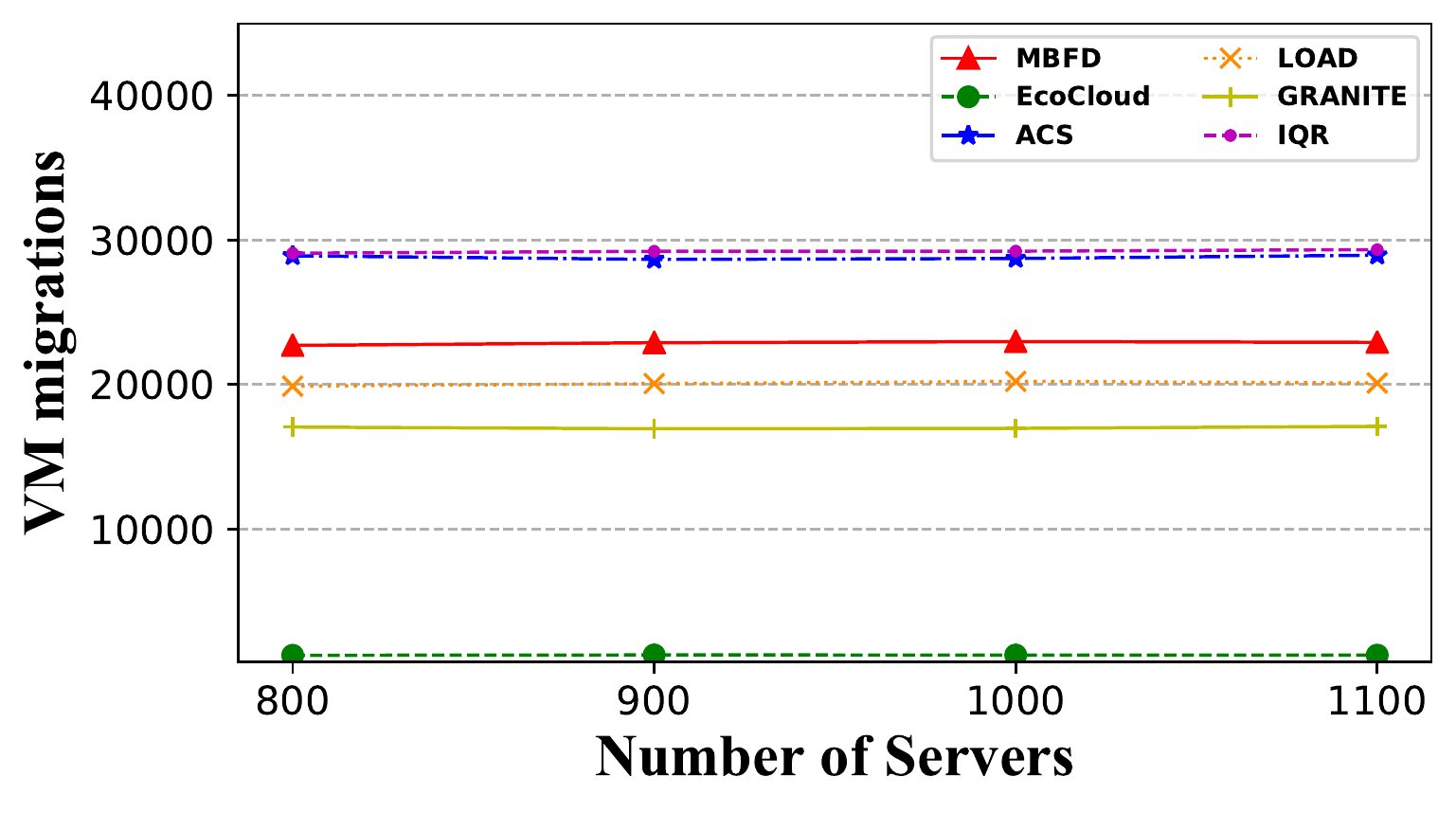}
		\caption{VM migrations}
		\label{fig:VMmigrationVaryingThresholdPlanetRatio}
	\end{subfigure}
	\begin{subfigure}{0.24\linewidth}
		\includegraphics[width=0.99\linewidth]{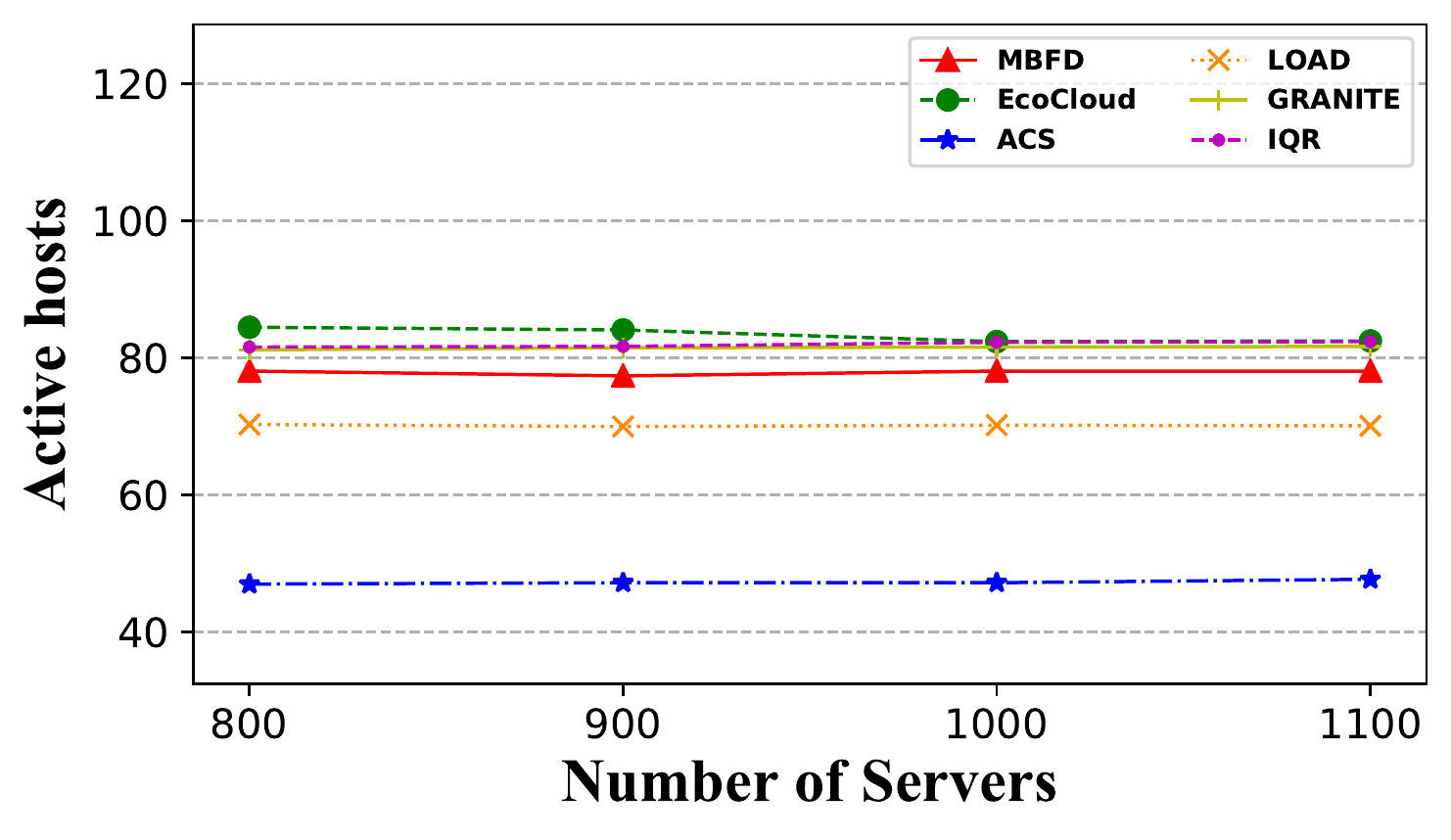}
		\caption{Number of active hosts}
		\label{fig:numberActiveHostVaryingThresholdPlanetRatio}
	\end{subfigure}
	\begin{subfigure}{0.24\linewidth}
		\includegraphics[width=0.99\linewidth]{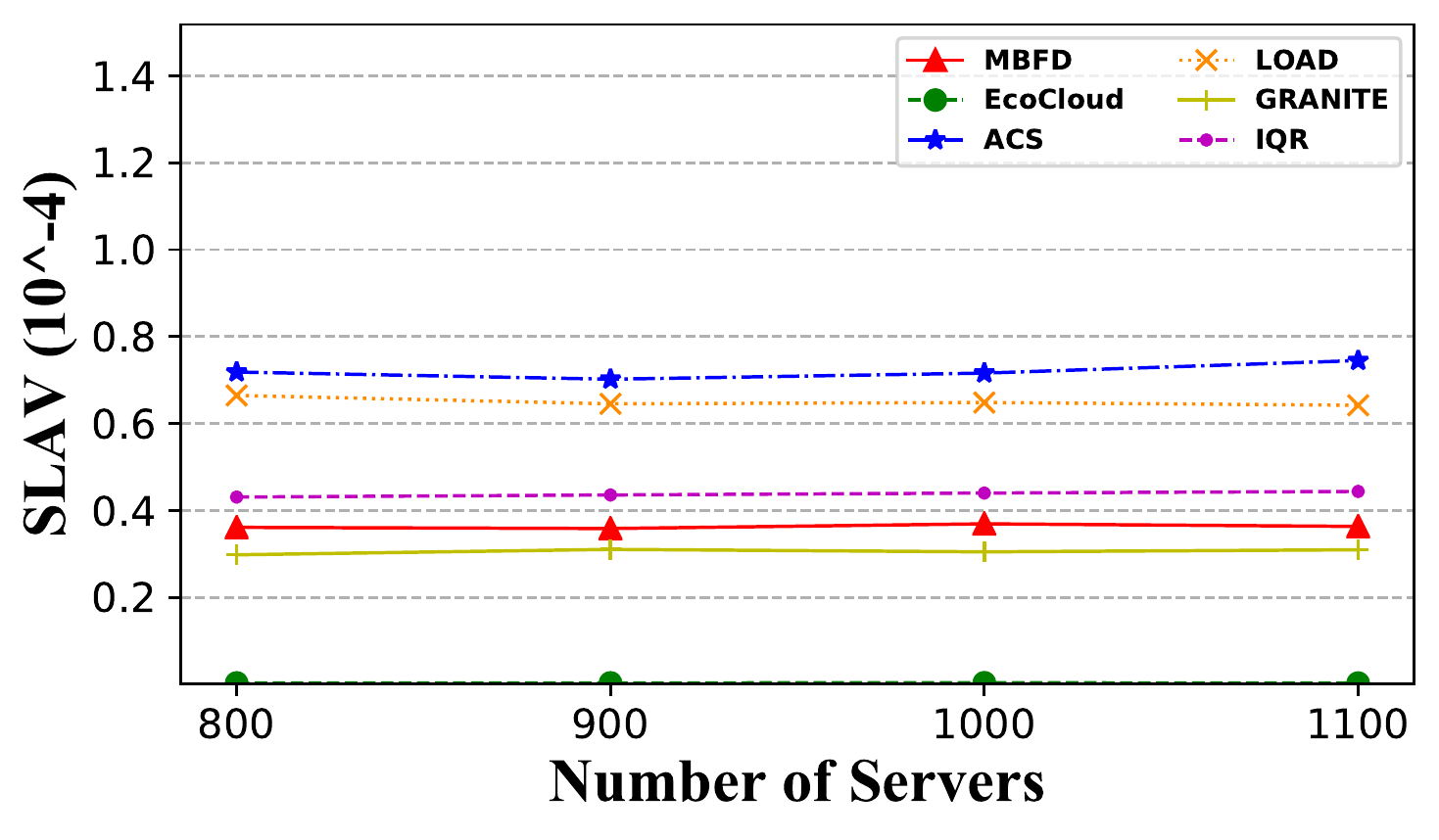}
		\caption{SLAV}
		\label{fig:SLAVVaryingThresholdPlanetRatio}
	\end{subfigure}
	
	\caption[VarPerOptCom]{Performance comparison of algorithms under PlanetLab workloads (varied number of servers with 800, 900, 1000 and 1100)}
	\label{fig:planetlabRatio}
	
\end{figure*}

In this section, to compare the performance of investigated algorithms, we conduct experiments for the five well-known and investigated algorithms based on different performance metrics and two traces. We also include one provided algorithm in CloudSim as baselines, the Interquartile Range (IQR) \cite{Calheiros2011a} that manages a dynamic threshold for overload detection. 

\subsection{Experiments Settings}

For host capacity, each host has two CPU cores with millions of instructions per second (MIPS) of 1880 or 2660, 4 GB of RAM and 1 TB of storage. We use the power model derived from HP ProLiant ML110 G4 or HP ProLiant ML110 G5, which has been used in \cite{beloglazov2012energy} and \cite{ACW2015}. According to the model, the energy consumption of the host with different utilization is shown in Table \ref{tab:power_model}. For VM configurations, we consider four types of VMs  with MIPS of 500, 1000, 1500 and 2500, and the VMs number of each type is randomly generated. The detailed specifications of hosts and VMs are shown in Table \ref{tab:hsotVMcapa}.

\begin{table}
	\centering
		 \caption{Host / VM Types and Capacity}
	\resizebox{0.48\textwidth}{!}{%
\label{tab:hsotVMcapa}
\begin{tabular}{|l|l|l|l|l|l|}

	\hline  Name & CPU (MIPS) & Cores  &  Memory & Bandwidth & Storage \\ 
	\hline
	\hline Host Type 1 & 1.86 GHz & 2 & 4 GB & 1 Gbit/s & 1 TB \\ 
	\hline Host Type 2 & 2.66 GHz & 2 & 4 GB & 1 Gbit/s & 1 TB  \\
			\hline
			\hline VM Type 1 & 2.5 GHz & 1 & 870 MB & 100 Mbit/s & 1 GB \\ 
			\hline VM Type 2 & 2.0 GHz & 1 & 1740 MB & 100 Mbit/s & 1 GB  \\
			\hline VM Type 3 & 1.0 GHz & 1 & 1740 MB & 100 Mbit/s & 1 GB  \\
			\hline VM Type 4 & 0.5 GHz & 1 & 613 MB & 100 Mbit/s & 1 GB  \\				
			\hline
 \end{tabular}
}
 \end{table}

We first carry out several experiments under {\color{black}synthetic} workloads, and then, to simulate the real cloud data center, we utilize the real workload data of the CoMon project provided by PlanetLab\cite{Park2006}. The workload includes CPU utilization data of thousands of VMs allocated to servers that are located in more than 500 places around the world. And the data is collected every five minutes for 10 days, which represents the workload in the real cloud environment. 

We select four metrics to evaluate the performance of these algorithms, including Energy Consumption, SLA violation percentage (SLAV), VM migration times and the number of active hosts. We choose these metrics as they have been adopted widely and also used in more than one algorithms as we discussed in Section V. Due to the page limitation, we evaluate SLAV to represent the SLA violations instead of average SLA violations. 

\subsection{Implementation Details}

The configurations in EcoCloud include a shape parameter $p$, $\alpha$ and $\beta$ in its probability function. We set $p = 3$, $\alpha = \beta = 0.25$ in our experiments, {\color{black} following the configurations of the original paper.} For LOAD, the reward and penalty parameters for updating learning automaton, $a$ and $b$, are both configured to 0.1. We utilize the parameters in the original paper of ACS, however, we don't have a training data set to estimate the initial pheromone level. Therefore, we set $M$ to be the total number of selected VMs, and $P$ is configured as the number of under-utilized servers. The configured parameters of MBFD and GRANITE are the same as the original paper.

\subsection{{\color{black}Synthetic} Workloads}
To show the performance under synthetic workloads, the first part of the evaluations focuses on the experiments under {\color{black}synthetic} workloads, which has been used in MBFD and ACS. 
We firstly generate the same number of hosts and VMs and vary the lower utilization threshold that defines when the host is under-utilized. And we varied the threshold from 0.1 to 0.5 with increment as 0.1. The higher utilization threshold is set as 0.4 more than the lower utilization threshold. {\color{black}We follow the configuration of the utilization threshold interval in \cite{beloglazov2012energy}}. Under each configuration, we randomly generate workloads to run the experiments and repeat 10 times. Both the number of hosts and VMs are set as 50.

Fig. \ref{fig:random1.0} shows the average results of the experiments under {\color{black}synthetic} workloads.
Based on the results, the higher values of lower utilization threshold can enable all the algorithms to achieve better energy efficiency, {\color{black}fewer VM migrations} and more SLAV. 
To be more specific, ACS performs the best on energy consumption by reducing 21.1\% power compared with MBFD. EcoCloud requires less than 600 migration times, which are much fewer than other algorithms. The reason that ACS achieves the best energy efficiency is that it has the lowest number of active hosts. For SLA violation comparison, these algorithms perform better when the lower threshold is set to be 0.1.

As setting the lower utilization threshold with 0.5 can achieve the best performance for all the algorithms, we fix the lower utilization threshold as 0.5, and run the experiments repeatedly with 10 times to show the variance of results{\color{black}, which are shown in Fig. \ref{fig:boxplotRandom1.0}.} We can notice that ACS can achieve the best performance in energy consumption with 31.2 kWh, and EcoCloud can reduce the VM migrations as 420.5 in average. 

To investigate the impacts of different numbers of hosts and VMs, we also fix the lower utilization threshold as 0.5, but configure the ratio of the number of servers to the number of VMs to be 1:1, 1:1.25, 1:1.5 and 1:1.75 respectively. With each ratio, the experiments are repeated 10 times, and the results as shown in  Fig \ref{fig:randomRatio}. We can observe that energy consumption increases significantly when there are more VMs. ACS is the most energy efficient algorithm and consumes 31.2 KWh and 31.5 kWh when the ratio is 1:1 and 1:1.75 respectively, which is about 21.2-34.4\% less compared to MBFD. EcoCloud achieves the best results in VM migrations. Although the investigated algorithms can reduce more energy than IQR, more SLAV happens in these algorithms.

\subsection{PlanetLab Workloads}
To show the algorithm performance under realistic traces, we also conduct experiments under PlanetLab workloads. 
We vary the lower CPU utilization threshold varies from 0.1 to 0.5, and the interval between the lower threshold and the higher threshold is fixed at 0.4. The number of hosts is configured as 800 and the number of VMs is retrieved from PlanetLab traces. The average results are shown in Fig. \ref{fig:PerformanceVaryingThresholdPlanetlab} by running 10 times of experiments, and each experiment is with a one-day PlanetLab trace. 

Fig. \ref{fig:energyConsumptionVaryingThresholdPlanet} shows the energy consumption comparison, in which the power {\color{black}decreases} when having a larger value of the lower utilization threshold. ACS consumes the least energy compared with other algorithms with 148.7 kWh when the lower utilization threshold is 0.4. EcoCloud consumes more energy consumption than MBFD because it keeps more servers to be active. 
The numbers of VM migrations are compared in Fig. \ref{fig:VMmigrationVaryingThresholdPlanet}. EcoCloud achieves an apparent reduction in this metric and only needs 1251 migrations on average when the lower threshold is set to be 0.5. LOAD achieves improvement in the number of migrations and reduces 12.4\% migrations compared with MBFD. GRANITE acquires better results when the lower utilization threshold increases. 
The number of active hosts comparison is shown in Fig. \ref{fig:numberActiveHostVaryingThresholdPlanet}, and ACS can shut down the maximum number of hosts.
Fig. \ref{fig:SLAVVaryingThresholdPlanet} shows the SLA violation percentage comparison, and the SLA violation percentage increases as the lower utilization threshold increases. As the figure shows, LOAD and ACS perform worse on this metric compared with MBFD and EcoCloud. EcoCloud maintains a low SLA violation percentage and thus ensures the quality of services.

We set the lower utilization threshold as 0.5 and the higher utilization threshold as 0.9, and run 10 times experiments repeatedly to show the variance of performance results as shown in Fig. \ref{fig:PerformanceVaryingThreshold_boxplot_Planetlab}. 
In Fig. {\ref{fig:boxplotEnergyConsumptionPlanet}}, we can notice that ACS and LOAD perform better in energy consumption compared with other baselines. The average energy consumption of ACS is 125.1 KWh while the MBFD, EcoCloud, and GRANITE consume more than 180 KWh. 
Fig. {\ref{fig:boxplotActiveHostPlanet}} shows the average results of the number of active hosts, and ACS can achieve the best results with 48 in average.
Fig. {\ref{fig:boxplotVMmigrationPlanet}} demonstrates the comparison of VM migrations, and EcoCloud focuses on optimizing this metric and reduces the number of VM migration to be under lower than 2000. Compared with MBFD and GRANITE, LOAD reduces the VM migrations. 
The SLA violation percentage comparison is presented in Fig {\ref{fig:boxplotSLAVPlanet}}, and  EcoCloud ensures SLA in a the best manner with $0.01\times 10^{-4} $. Although ACS saves more energy, it performs worst in reducing SLA violations with $0.71\times 10^{-4} $.

To evaluate the performance with a different number of hosts, We also vary the number of hosts from 800 to 1100 in our experiments as shown in Fig. \ref{fig:planetlabRatio}. As the number of hosts increases, we can notice that ACS always consumes the minimum power, and the least number of VM migration happens in EcoCloud. ACS only keeps 48 active hosts and leads to the maximum SLAV as $0.72\times 10^{-4} $.

In conclusion, we can notice that under both workloads, ACS can achieve the best energy efficiency in most cases as it also has the least number of active hosts. EcoCloud can reduce more VM migrations compared with other algorithms and trigger the least SLA violations under PlanetLab workloads. We can notice that meta-heuristic algorithms, e.g. ACS, can achieve better performance in energy compared with heuristic algorithms as it searches larger solution space. As the pioneer of VM consolidation-based energy efficient algorithm for cloud data centers, MBFD has been widely accepted as it is easy to implement and efficient. Although the performance of MBFD has been outperformed by recent algorithms, the main idea of MBFD has been referred, such as in GRANITE and LOAD, where the holistic energy and SLA violation are optimized respectively. GRANITE is suggested for the case that aims to optimize more energy-consuming components rather than only the hosts. EcoCloud is suitable for the scenarios that VM migrations should be minimized, such as the network is the bottleneck of the whole system, as it can reduce VM migrations significantly. LOAD can achieve good performance if the future resource usage can be accurately predicted, thus it suits the conditions that the system has adequate history resource usage data or resource usage has a typical pattern, like Wikipedia traces.

\section{Conclusions and Future Work}

In this paper, we present an investigation of 5 state-of-the-art energy efficient algorithms based on VM consolidations for cloud data centers. We discuss and compare the algorithms from multiple perspectives, including core ideas, architectures, modelling and algorithm complexity. We also implement these algorithms in CloudSim and configure the parameters as noted in these algorithms. By conducting experiments under both {\color{black}synthetic} and real traces, the results show that these algorithms can efficiently reduce the energy consumption while balancing the trade-offs between energy and some other metrics, such as VM migrations and SLA violations. 

Based on our investigation, some future research directions are identified as follows:
\begin{itemize}
    \item Utilization threshold settings have important impacts on energy consumption, thus self-adaptive threshold configuration approach can be further investigated. 
    \item Most VM consolidation-based energy efficient algorithms consider the CPU resource as the bottleneck and the key factor in their power models. Resource usage and energy consumption related to other resources, e.g. network, can be further considered.  
    \item Meta-heuristic algorithms have shown their better performance in improving energy consumption compared with traditional heuristic algorithms, however, the related time cost should be reduced, which can be achieved by reducing the solution space. 
    \item Evaluations can be conducted under recently published workloads, such as Google traces and Alibaba traces. 
\end{itemize}
\section*{Acknowledgment}

This work is supported by the SIAT Innovation Program for Excellent Young Researchers (No. 55Y05504).

\bibliographystyle{IEEEtran}
\bibliography{library}

\end{document}